\documentclass[trackchanges]{aastex701}

\shorttitle{A Multiscale Decomposition Strategy with Wavelet Transform for F10.7 Prediction}
\shortauthors{Ma et al.}
\usepackage{multirow} 
\usepackage{longtable}
\usepackage{array}
\usepackage{booktabs}
\usepackage{ltablex}
\usepackage{xcolor}
\usepackage{float}
\usepackage{CJKutf8}
\usepackage{graphicx}
\usepackage{array}
\usepackage{url}
\usepackage{xltabular}
\usepackage{rotating}
\usepackage{makecell}
\usepackage{xcolor}
\begin{document}

\title{F10.7 Index Prediction: A Multiscale Decomposition Strategy with Wavelet Transform for Performance Optimization}

\correspondingauthor{Xuebao Li}
\email{lixuebaozhengyanfang@gmail.com}  

\correspondingauthor{Yanfang Zheng}
\email{zyf062856@163.com}

\author[0009-0005-0190-9261]{Xuran Ma}
\affiliation{School of Computer Science, Jiangsu University of Science and Technology, Zhenjiang 212100, People$^{\prime}$s Republic of China}
\email{15252132432@163.com}

\author[0000-0003-0397-4372]{Xuebao Li}
\affiliation{School of Computer Science, Jiangsu University of Science and Technology, Zhenjiang 212100, People$^{\prime}$s Republic of China}
\email{lixuebaozhengyanfang@gmail.com}  

\author[0000-0003-0229-3989]{Yanfang Zheng}
\affiliation{School of Computer Science, Jiangsu University of Science and Technology, Zhenjiang 212100, People$^{\prime}$s Republic of China}
\email{zyf062856@163.com}

\author[0009-0004-9110-0425]{Yongshang Lv}
\affiliation{School of Computer Science, Jiangsu University of Science and Technology, Zhenjiang 212100, People$^{\prime}$s Republic of China}
\email{sccy0329@163.com}

\author[0009-0004-6402-7509]{Xiaojia Ji}
\affiliation{School of Computer Science, Jiangsu University of Science and Technology, Zhenjiang 212100, People$^{\prime}$s Republic of China}
\email{1092028768@qq.com}

\author[0009-0001-6084-5430]{Jiancheng Xu}
\affiliation{School of Computer Science, Jiangsu University of Science and Technology, Zhenjiang 212100, People$^{\prime}$s Republic of China}
\email{19816392728@163.com}

\author[0009-0009-9722-5794]{Hongwei Ye}
\affiliation{School of Computer Science, Jiangsu University of Science and Technology, Zhenjiang 212100, People$^{\prime}$s Republic of China}
\email{hongweiye777@163.com}

\author[0009-0005-2367-9647]{Zixian Wu}
\affiliation{School of Computer Science, Jiangsu University of Science and Technology, Zhenjiang 212100, People$^{\prime}$s Republic of China}
\email{3041824536@qq.com}

\author[0000-0001-7460-3275]{Shuainan Yan}
\affiliation{State Key Laboratory of Space Weather, National Space Science Center, Chinese Academy of Sciences, Beijing 100190, People$^{\prime}$s Republic of China}
\email{shuainanyan2021@163.com}

\author[0000-0002-4749-623X]{Liang Dong}
\affiliation{Yunnan Astronomical Observatory, Chinese Academy of Sciences, Kunming 650216, China}
\email{dongliang@ynao.ac.cn}

\author[0000-0002-7149-0997]{Zamri Zainal Abidin}
\affiliation{Radio Cosmology Lab, Centre for Astronomy and Astrophysics Research, Department of Physics, Faculty of Science, Universiti Malaya, 50603 Kuala Lumpur, Malaysia}
\affiliation{National Centre for Particle Physics, Universiti Malaya, 50603 Kuala Lumpur, Malaysia}
\email{zzaa@um.edu.my}

\author[0009-0008-1279-1870]{Xusheng Huang}
\affiliation{School of Computer Science, Jiangsu University of Science and Technology, Zhenjiang 212100, People$^{\prime}$s Republic of China}
\email{1436106554@qq.com}

\author[0009-0003-0325-1366]{Shunhuang Zhang}
\affiliation{School of Computer Science, Jiangsu University of Science and Technology, Zhenjiang 212100, People$^{\prime}$s Republic of China}
\email{zhangshunhuang@icloud.com}

\author[0009-0006-5294-3571]{Honglei Jin}
\affiliation{School of Computer Science, Jiangsu University of Science and Technology, Zhenjiang 212100, People$^{\prime}$s Republic of China}
\email{hongleijin53@gmail.com}

\author[0000-0001-5130-9997]{tarik abdul latef}
\affiliation{Department of Electrical Engineering, Faculty of Engineering, Universiti Malaya, 50603 Kuala Lumpur, Malaysia}
\email{tariqlatef@um.edu.my}

\author[0000-0002-1320-8711]{Noraisyah Mohamed Shah}
\affiliation{Department of Electrical Engineering, Faculty of Engineering, Universiti Malaya, 50603 Kuala Lumpur, Malaysia}  
\email{noraisyah@um.edu.my}

\author[0000-0002-0563-5000]{Mohamadariff Othman}
\affiliation{Department of Electrical Engineering, Faculty of Engineering, Universiti Malaya, 50603 Kuala Lumpur, Malaysia} 
\affiliation{National Centre for Particle Physics, Universiti Malaya, 50603 Kuala Lumpur, Malaysia} 
\email{mohamadariff@um.edu.my}

\author[0000-0003-4003-4793]{Kamarul Ariffin Noordin}
\affiliation{Department of Electrical Engineering, Faculty of Engineering, Universiti Malaya, 50603 Kuala Lumpur, Malaysia}
\email{kaharudin@um.edu.my}

\begin{abstract}

In this study, we construct Dataset A for training, validation, and testing, and Dataset B to evaluate generalization. We propose a novel F10.7 index forecasting method using wavelet decomposition, which feeds F10.7 together with its decomposed approximate and detail signals into the iTransformer model. We also incorporate the International Sunspot Number (ISN) and its wavelet-decomposed signals to assess their influence on prediction performance. Our optimal method is then compared with the latest method from \citet{yan2025research} and three operational models (SWPC, BGS, CLS). Additionally, we transfer our method to the PatchTST model used in \citet{YE20246309} and compare our method with theirs on Dataset B. Key findings include: (1) The wavelet-based combination methods overall outperform the baseline using only F10.7 index. The prediction performance improves as higher-level approximate and detail signals are incrementally added. The Combination 6 method—integrating F10.7 with its first to fifth level approximate and detail signals—outperforms methods using only approximate or detail signals. (2) Incorporating ISN and its wavelet-decomposed signals does not enhance prediction performance. (3) The Combination 6 method significantly surpasses \citet{yan2025research} and three operational models, with RMSE, MAE, and MAPE reduced by 18.22\%, 15.09\%, and 8.57\%, respectively, against the former method. It also excels across four different conditions of solar activity. (4) Our method demonstrates superior generalization and prediction capability over the method of \citet{YE20246309} across all forecast horizons. To our knowledge, this is the first application of wavelet decomposition in F10.7 prediction, substantially improving forecast performance.

\end{abstract}


\section{Introduction}\label{sec:Introduction}
Solar activity is the core physical process driving changes in the Sun-Earth space environment. It profoundly impacts space weather at earth \citep{baker2013major}, radio communications \citep{lanzerotti2004solar}, and even global climate \citep{bard2006climate} through energy releases in forms such as solar flares and coronal mass ejections \citep{pick2008sixty}. \citet{covington1969solar} first proposed the concept of the solar 10.7 cm radio flux (F10.7 index), a physical quantity formed by the superposition of radiation from all active regions on the solar disk. This index is largely unaffected by atmosphere and diurnal variations of earth, and it is highly sensitive to changes in solar active region conditions \citep{tapping201310}. Therefore, accurate prediction of the F10.7 index is crucial for the operational safety of satellites and technical systems such as communication and navigation.

In previous studies, many researchers commonly used statistical methods to predict the F10.7 index \citep[e.g.,][]{warren2017linear,wang2018linear,zhao2008historical}. \citet{warren2017linear} proposed a linear prediction model using the observed F10.7 index from the past 81 days to forecast the F10.7 index for the next 45 days. The British Geological Survey (BGS) generated 27 day forecasts of the daily F10.7 index based on the autoregressive integrated moving average (ARIMA) model. \citet{si2010modeling} employed an autoregressive method for 27 day ahead forecasts, which demonstrated high accuracy and effectiveness only during periods of low solar activity when the 27 day periodicity of F10.7 was evident. The Space Weather Prediction Center (SWPC) used the least squares method to fit nonlinear curves to predict future F10.7 levels \citep{miesch2025solar}. Since the F10.7 index correlates with various solar activity phenomena, researchers often incorporated related solar observation data, such as extreme ultraviolet (EUV) images and active region areas, to achieve more accurate predictions. \citet{lei2019mid} proposed a medium-term prediction method based on EUV images, defining the PSR index to construct an empirical model that improved the prediction accuracy of the F10.7 index for the next 27 days. \citet{qian2019f} developed a forecasting method for the F10.7 index based on statistical analysis of active region areas on the solar disk, achieving predictions by classifying active region areas and establishing forecasting formulas. The above methods provide a basic solution for predicting the F10.7 index, but they struggle to adapt to sudden changes in solar activity, thus limiting further performance improvements.

In recent years, machine learning, particularly deep learning, has been widely applied to space weather forecasting \citep{yan2024application,YE20246309,stevenson2022deep,zhang2024forecasting}. The Collect Localization Satellites (CLS) combined multiple different solar radio flux time series with neural network models to predict the F10.7 index for the next 30 days \citep{yaya2017solar}. \citet{yan2024application} constructed a BiLSTM-Attention model and proposed a conversion average calibration method to preprocess F10.7 index observed by Long and Short Wave Solar Precision Flux Radio (L\&S) telescope in Langfang, China, achieving higher prediction accuracy and stability. \citet{stevenson2022deep} employed the N-BEATS model for the prediction of F10.7 index, a novel univariate deep residual structure based on feedforward neural networks, which outperformed baseline methods. \citet{YE20246309} first applied the Transformer-based PatchTST model to the prediction of the F10.7 index, improving the accuracy of the prediction and demonstrating superior adaptability in model uncertainty. \citet{zhang2024forecasting} used the Transformer-based Informer model to predict the F10.7 index, showing better performance and accuracy compared to other commonly used prediction techniques, particularly during the solar activity descending phase and at the solar activity minimum.

Although some progress has been made in research on F10.7 index prediction based on deep learning models, its prediction performance still needs to be improved. Therefore, researchers have introduced signal decomposition processing techniques into the prediction of F10.7 index. \citet{luo2021new} utilized empirical mode decomposition (EMD) to decompose the F10.7 index into multiple components with different frequency characteristics, then employed LSTM model to predict each component separately, and finally obtained the final F10.7 prediction value through information reconstruction. \citet{marcucci2023deep} performed fast iterative filtering (FIF) decomposition on multiple different solar radio flux time series. On this basis, they improved the prediction performance by inputting the original F10.7 index and the intrinsic mode components highly correlated with F10.7 into the LSTM model. \citet{hao2024f10} proposed a LSTM combined with variational mode decomposition (VMD) method which exhibits strong predictive capability for the F10.7 index during solar cycle 24. They employed VMD to decompose the F10.7 index into several intrinsic mode functions (IMF), and then utilized LSTM model to forecast each IMF separately. Finally, they aggregated all the prediction results to obtain the final F10.7 prediction value. \citet{yan2025research} employed a factor decomposition method that uses a Savitzky-Golay (SG) filter to separate the F10.7 index into smooth and jitter components, then predicted each component separately with the iTransformer model and combined the results to enhance prediction performance. Additionally, they introduced other features such as soft X-ray flare index, magnetic type of the active region, and X-ray background flux to enhance understanding of the potential physical processes of solar activity. The above research fully demonstrated that the signal decomposition method can effectively extract different characteristic components from the F10.7 index. The wavelet decomposition can precisely capture the multi-scale features of the data and has certain advantages in processing periodic time series \citep{rhif2019wavelet}. The F10.7 index is a quasi-periodic time series \citep{roy2019search}. However, from the existing literature, there has been no research on the prediction of the F10.7 index based on wavelet decomposition.

In our work, we construct Dataset A for model training, validation, and testing, and Dataset B for evaluating generalization performance. We present a novel F10.7 index forecasting method based on wavelet decomposition. The original index, along with its decomposed approximate and detail components, serves as the input to the iTransformer model. To investigate the influence of other features closely related to the F10.7 index, and informed by prior research \citep{henney2012forecasting} on the predictive utility of the International Sunspot Number (ISN), we also incorporate the ISN and its corresponding approximate and detail signals as additional model inputs. We conduct a comparative study to evaluate the prediction performance of our proposed method against that of \citet{yan2025research} and three operational models from SWPC, BGS, and CLS during the same time period. Furthermore, we extend our method to the PatchTST model employed by \citet{YE20246309} and compare our method with their forecasting method on Dataset B. The structure of this paper is arranged as follows: Section \ref{sec:Data} introduces the data, Section \ref{sec:Method} describes the method, Section \ref{sec:Results} presents the results, and Section \ref{sec:Conclusions} presents the conclusions and discussions.

\section{Data}\label{sec:Data} 
In this study, we construct training, validation, and testing datasets based on the F10.7 index, its wavelet-decomposed signals, the ISN, and the wavelet-decomposed signals of ISN. The daily F10.7 index is measured by the Dominion Radio Astrophysical Observatory (DRAO) in Canada. To evaluate the generalization performance of the model, we incorporate daily F10.7 index from the L\&S Telescope in Langfang, China, as well as contemporaneous data from DRAO. In our work, the time resolution for the F10.7 index is one day.
\subsection{Training, validation, and testing datasets}\label{subsec:datasetA}
\begin{figure}[ht]
\centering
\includegraphics[width=0.85\linewidth]{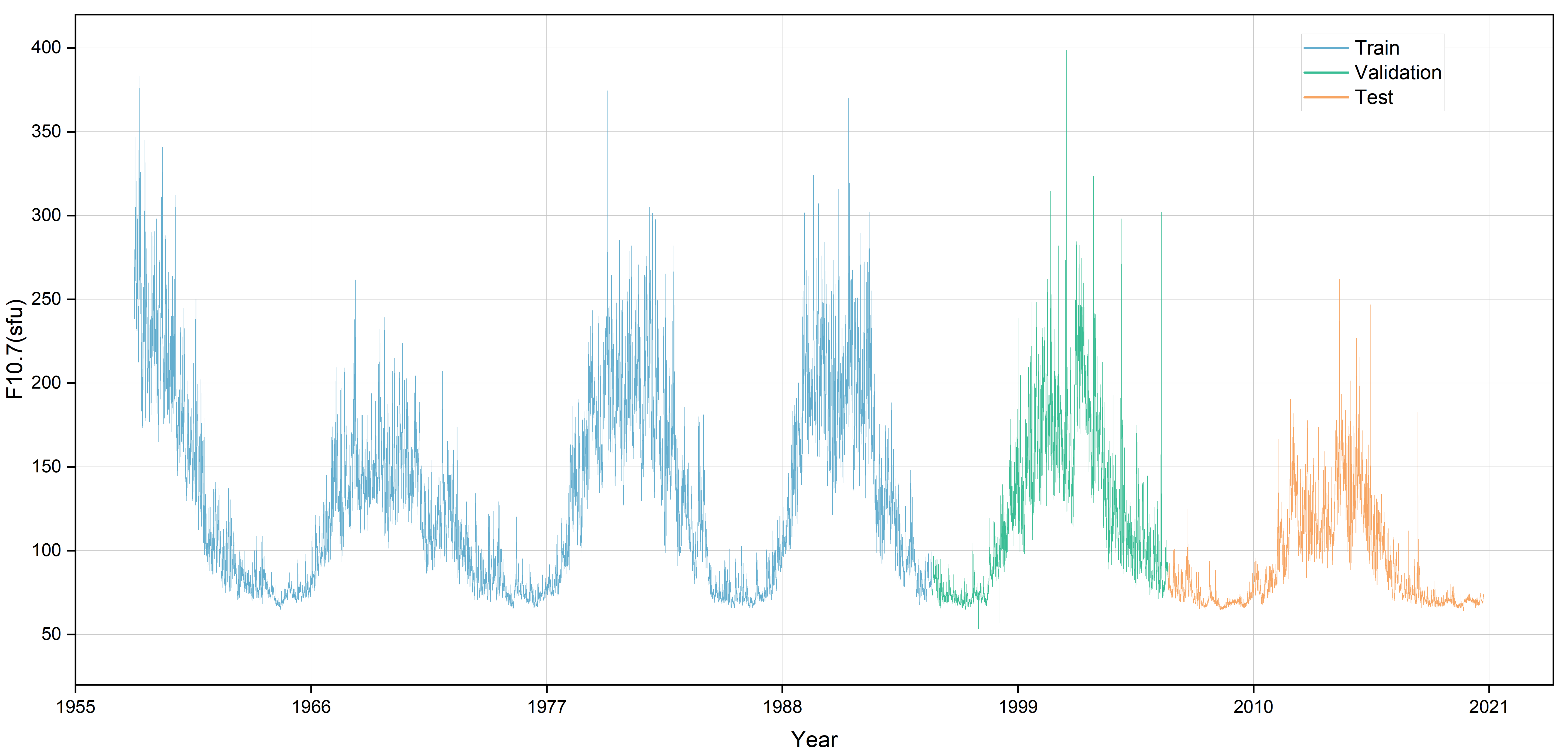}
\caption{Daily F10.7 index of DRAO data from 1957 to 2020. The training set, validation set and testing set are divided according to solar cycles. The blue part in the data is used for model training, the green part for model validation, and the yellow part for model testing.}
\label{fig1}
\end{figure}

We obtain the daily F10.7 index and ISN data from the space weather monitoring system at the DRAO in Canada (\url{http://www.celestrak.com/SpaceData/SW-All.txt}). To avoid introducing future information into the model training, we divide the F10.7 index and ISN into training, validation, and testing datasets according to the solar cycle, and perform wavelet decomposition on them respectively, which are recorded as Dataset A. The specific details of Dataset A are described in Section \ref{subsec:result1}. The training dataset covers data from 1957 to 1994, the validation dataset covers data from 1995 to 2005, and the testing dataset covers data from 2006 to 2020. Figure \ref{fig1} illustrates the distribution of the daily F10.7 index in Dataset A.
\subsection{Dataset for generalization performance evaluation}\label{subsec:datasetB}
\begin{figure}[ht]
	\centering
	\includegraphics[width=0.9\linewidth]{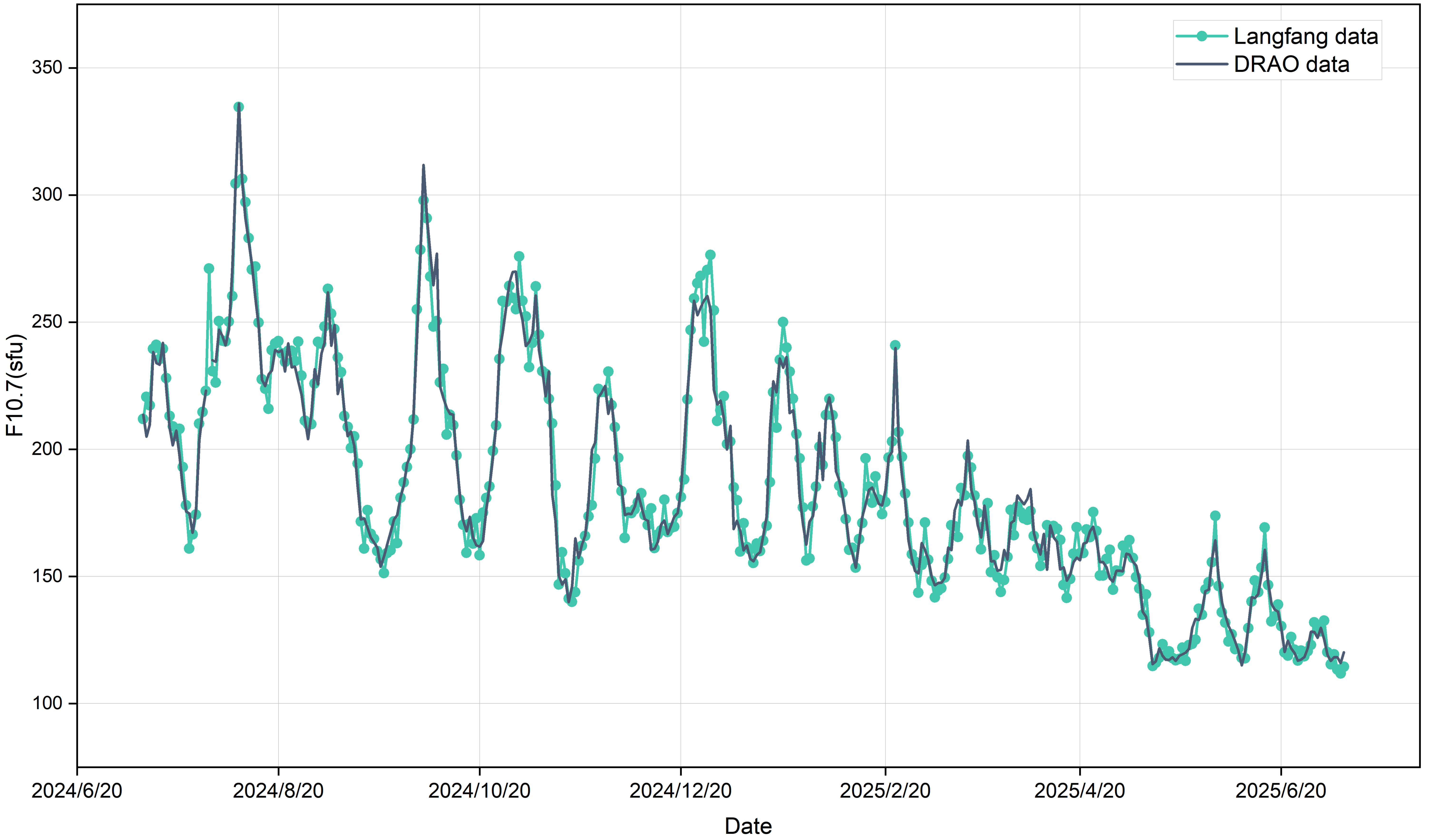}
	\caption{Comparison of Langfang data and DRAO data for the time period 2024-2025.}
	\label{fig2}
\end{figure}

To validate whether our model maintains good prediction performance across different telescope observations during the same time period, we construct the dataset for generalization performance evaluation, referred to as Dataset B. This dataset integrates concurrent F10.7 index from both the DRAO and the L\&S Telescope in Langfang, China \citep{yan2024application,YE20246309}, along with their corresponding wavelet-decomposed signals. The newly added data span from July 10, 2024, to July 9, 2025, as shown in Figure \ref{fig2}. The Langfang data exhibit distribution patterns and trends that are generally similar to those of the DRAO data during the same observation period.

\section{Method}\label{sec:Method} 
\subsection{Discrete wavelet decomposition}\label{subsec:met1}

\begin{figure}[H]
	\centering
	\includegraphics[width=0.4\linewidth]{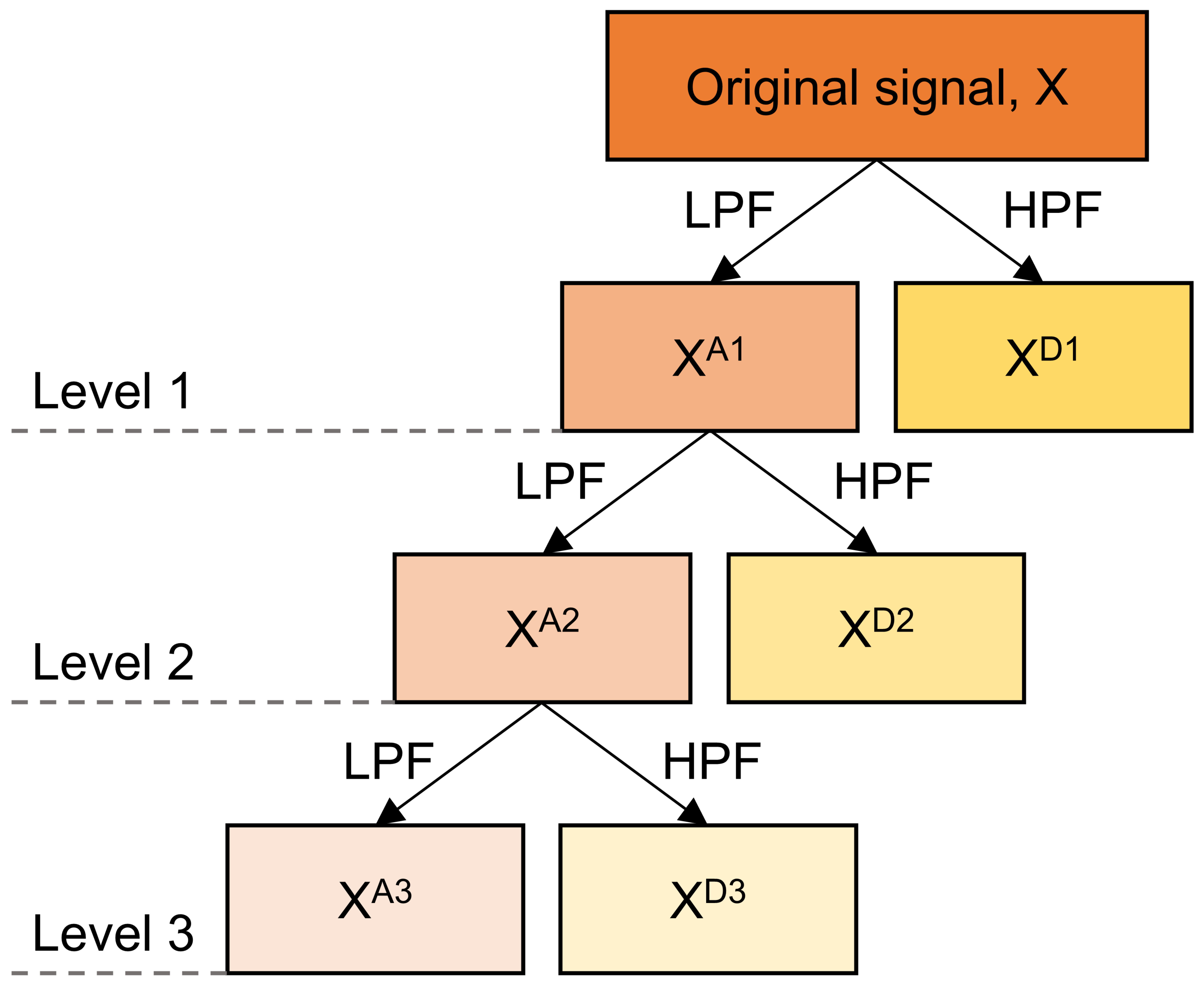}
	\caption{Wavelet multi-level decomposition. X$^\mathrm{AL}$ and X$^\mathrm{DL}$ respectively represent the approximate signals and the detail signals of the L-th level.}
	\label{fig3}
\end{figure}

Signal processing techniques, such as wavelet transform and fourier transform, have been successfully integrated with deep learning models to improve the accuracy of tasks in ocean wave prediction, semantic segmentation, and other fields \citep{wang2024hybrid,zhang2024segcft}. The choice of different wavelet bases directly affect the quality of feature extraction and the final prediction performance. In this study, we employ the discrete bior2.2 wavelet for feature enhancement. This wavelet possesses both symmetry and linear phase characteristics, which can effectively reduce phase distortion during the signal decomposition \citep{Burrus1997IntroductionTW}. It is likely suitable for processing non-stationary time series, such as the F10.7 index and ISN. Figure \ref{fig3} illustrates the process of wavelet multi-level decomposition. As shown in Figure \ref{fig3}, the process of discrete wavelet decomposition separates the original signal into an approximate signal and a detail signal using a low-pass filter (LPF) and a high-pass filter (HPF), respectively \citep{ocak2009automatic}. Following the decomposition process illustrated in Figure \ref{fig3}, we define the L-th level approximate signal of the F10.7 index as F10.7$^\mathrm{AL}$ and its detail signal as F10.7$^\mathrm{DL}$. Correspondingly, we define the L-th level approximate signal of the ISN as ISN$^\mathrm{AL}$ and its detail signal as ISN$^\mathrm{DL}$.

\subsection{iTransformer}\label{sec:met2} 
\begin{figure}[ht]
	\centering
	\includegraphics[width=0.8\linewidth]{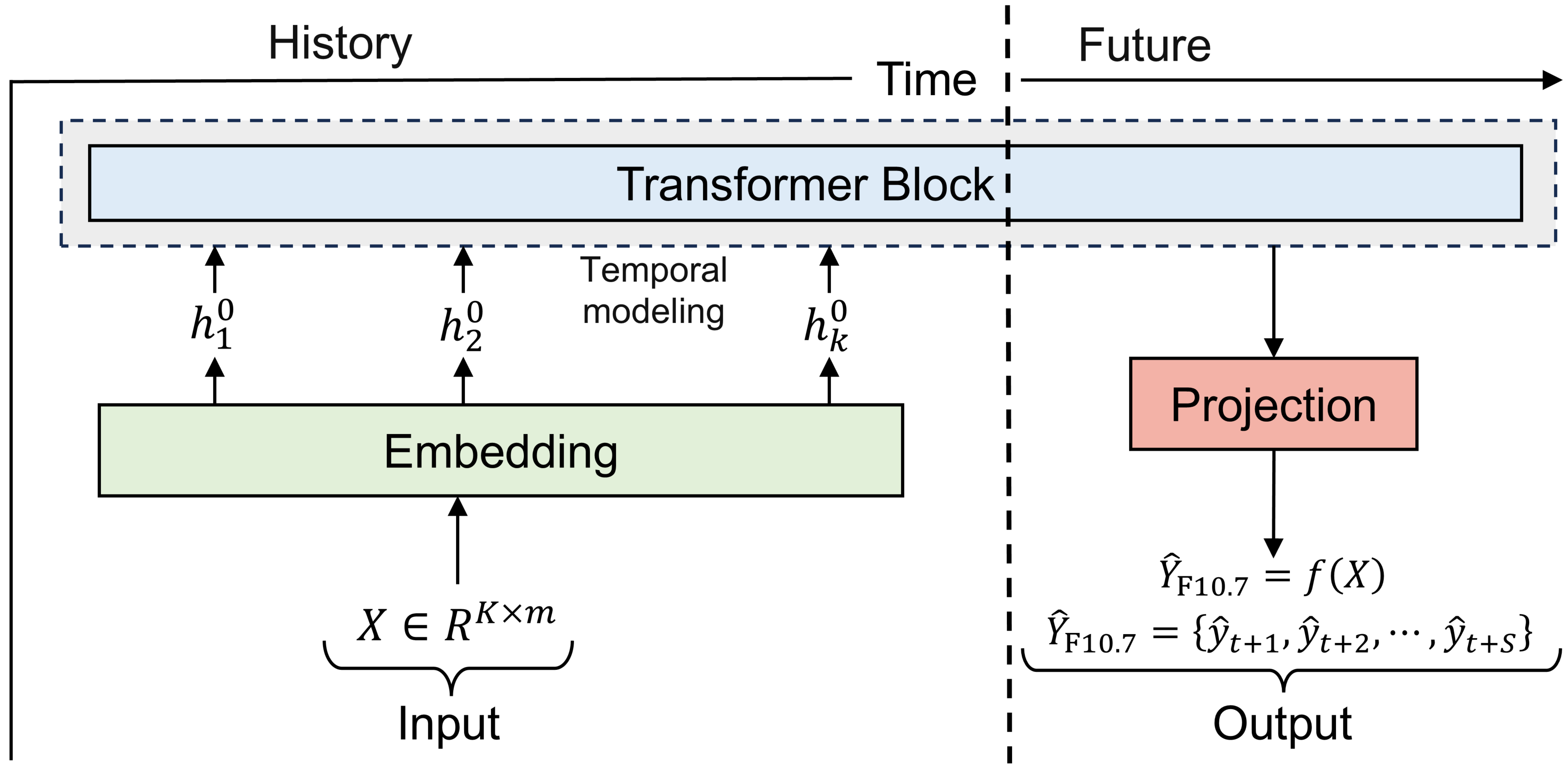}
	\caption{The prediction scheme of the F10.7 index for future periods based on the iTransformer model.}
	\label{fig4}
\end{figure}

\begin{figure}[ht]
	\centering
	\includegraphics[width=0.9\linewidth]{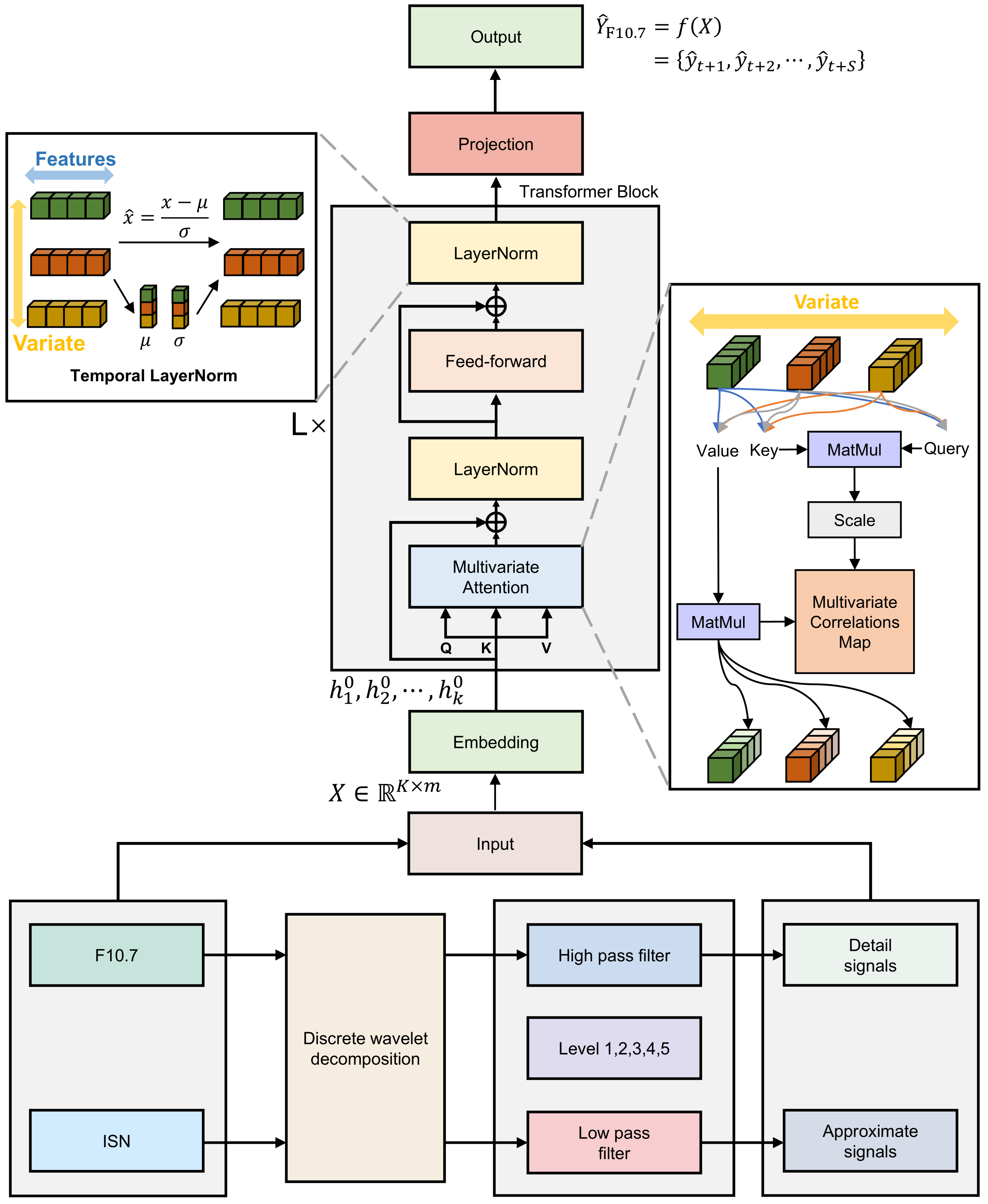}
	\caption{The schematic diagram for F10.7 prediction method based on the iTransformer model and wavelet decomposition. The lower section of the diagram displays the integrated inputs to the model, which include the F10.7 index, ISN, their respective approximate signals and detail signals derived from wavelet decomposition, along with the original data. The upper section presents the iTransformer model architecture, which primarily consists of an embedding layer, a projection layer, and multiple inverted Transformer Blocks.}
	\label{fig5}
\end{figure}

The iTransformer model, proposed by \citet{liu2024itransformer}, represents a novel time series forecasting model based on the Transformer framework. Unlike Transformer models which model sequences along the temporal dimension, the core innovation of the iTransformer lies in reversing the dimension of attention mechanism application. It treats the entire sequence of each variable as a token and performs self-attention calculations along the variable dimension in multivariate time-series forecasting tasks. Furthermore, this model processes each individual time series variable independently, effectively capturing interdependencies among variables, making it suitable for future period prediction tasks. In this study, we focus on a set of multivariate time series comprising F10.7, F10.7$^\mathrm{AL}$, F10.7$^\mathrm{DL}$, ISN, ISN$^\mathrm{AL}$, and ISN$^\mathrm{DL}$. We design various input combinations to investigate their impact on the forecasting performance for future F10.7 periods. Figure \ref{fig4} illustrates the prediction scheme of the F10.7 index for future periods based on the iTransformer model. Figure \ref{fig5} illustrates the schematic diagram for F10.7 prediction method based on the iTransformer model and wavelet decomposition. The architecture of model includes the embedding, projection, and Transformer blocks, as shown in Figure \ref{fig5}.

We define the input variable combination as $V \subseteq \{ \text{F10.7, F10.7}^{\mathrm{AL}}, \text{F10.7}^{\mathrm{DL}}, \text{ISN, ISN}^{\mathrm{AL}}, \text{ISN}^{\mathrm{DL}} \}$, where $|V| = K$ represents the number of variables. The input data forms a historical time series matrix $X \in \mathbb{R}^{K \times m}$, where $m$ represents the length of the historical time series, that is lookback window. For the $k$-th variable ($k = 1, 2, \dots, K$), the historical time series, $X_{k} = \{x_{t-m+1,k}, x_{t-m+2,k}, \cdots, x_{t,k}\}$, is input into the model, where $t$ represents the starting time point for prediction. Our objective is to predict the F10.7 for the future $S$-days period,  $\hat{Y}_{\text{F10.7}} = \{\hat{y}_{t+1}, \hat{y}_{t+2}, \cdots, \hat{y}_{t+S}\} \in \mathbb{R}^{1 \times S}$, which is a sequence with a prediction length of $S$ days. The relationship among the model, historical time series matrix, and the predicted sequence F10.7 can be defined as:
\begin{equation}
	\hat{Y}_{\mathrm{F}10.7} = f(X).
\end{equation}
Where, $f$ represents the iTransformer model, and the specific prediction modules of the model are as follows:

\textbf{1) Embedding:} For different input combinations $V$, the model first processes each input sequence $X_k$ through embedding. It transforms the sequences into high-dimensional feature tokens, preparing the data for subsequent operations. We formalize the embedding process as follows:
\begin{equation}
	H^{0} = \text{Embedding}\left( \left\{ X_k, k=1,2,\cdots,K \right\} \right).
\end{equation}
Where, $ H^{0} = \left\{ h_{k}^{0} \mid k=1, \cdots, K \right\} \in \mathbb{R}^{K \times D} $ represents the output of the embedding layer, where $D$ denotes the feature dimension. Each token $h_k^0$ corresponds to the representation of the $k$-th variable.

\textbf{2)Transformer Block:} The core architecture of iTransformer consists of a stack of $L$ inverted Transformer Blocks. These blocks capture global dependencies among variables by operating on the token dimension. The computation at the $l$-th layer is defined as:
\begin{equation}
	H^{l} = \text{TransformerBlock}\left(H^{l-1}\right), \quad l=1, \cdots, L.
\end{equation}
Each Transformer Block consists of layer normalization, self-attention mechanism, and feed-forward network. The specific structure is as follows:

\textbf{Layer normalization.}Treating each variable as an independent token, it performs layer normalization along the feature dimension for each variable, which preserves the representational independence between variables. We define the layer normalization as follows:
\begin{equation}
	\text{LayerNorm}(H) = \left\{ \frac{h_k - \text{Mean}(h_k)}{\sqrt{\text{Var}(h_k)}} \right\}, {k=1,\cdots,K}.
\end{equation}
Where, $H$ represents the input to the current inverted Transformer Block, which is the output $H^l$ from the previous layer. The functions \text{Mean} and \text{Var} calculate the mean and variance, respectively.

\textbf{Self-Attention Mechanisms.}The output from the layer normalization undergoes linear projections to generate the Query ($Q$), Key ($K$), and Value ($V$) matrices. The mechanism then calculates the attention weights as follows:
\begin{equation}
	\text{Attention}(Q,K,V) = \text{softmax}\left( \frac{QK^T}{\sqrt{d_k}} \right) V.
\end{equation}
Where, \text{softmax} represents the normalized exponential function, and $d_k$ is the dimensionality of the Key vectors.

\textbf{Feed-forward network.} The output from the self-attention mechanism first undergoes a residual connection with the input $H$ of the Transformer Block. The resulting sum then passes through another layer normalization, producing $x$. This normalized output $x$ is subsequently fed into the feed-forward network, enabling the model to learn more complex features and relationships. We define the feed-forward network as follows:
\begin{equation}
	\mathrm{FFN}(x) = \mathrm{Linear}_2(\mathrm{GELU}(\mathrm{Linear}_1(x))).
\end{equation}
Where, $\mathrm{Linear}_1$ projects the input vector x into a higher-dimensional feature space. The GELU (Gaussian Error Linear Unit) activation function then introduces non-linear transformations, allowing the model to learn more complex functional mappings. Finally, $\mathrm{Linear}_2$ projects the high-dimensional features back to the original model dimension, ensuring dimensional consistency with the main network trunk for subsequent operations.

\textbf{3)Projection:} After processing through $L$ layers of inverted Transformer Blocks, the final output $H^L$ passes through a projection layer. This layer maps the abstract features learned by the model back to the scale of the target sequence, directly generating the predicted F10.7 index values for the subsequent $S$ days, $\hat{Y}_{\text{F10.7}} = \{\hat{y}_{t+1}, \hat{y}_{t+2}, \cdots, \hat{y}_{t+S}\}$. We define this projection as:
\begin{equation}
	\hat{Y}_{\mathrm{F10.7}} = \mathrm{Projection} \left( H^{L} \right).
\end{equation}

\subsection{Model training and evaluation indicators}\label{subsec:training&evaluation}
When evaluating the prediction performance of the iTransformer model integrated with wavelet-decomposed signals for the F10.7 index, it is crucial to train the model by setting an optimal combination of hyperparameters. The detail configuration of the optimal hyperparameters is shown in Table \ref{tab1}. This study focuses on the prediction performance of the proposed method in short term (1 to 3 days) and short and medium-term (5 to 45 days) predictions of the F10.7 index. The prediction sequences represent forecasts for the future time periods of S days: 1, 2, 3, 5, 10, 15, 20, 25, 27, 30, 35, 40, and 45 days.
\begin{table}[H]
	\centering
	\caption{Hyperparameter settings of the iTransformer model that integrates wavelet-decomposed signals.}
	\label{tab1}
	\begin{tabular}{cc}
		\toprule
		\textbf{Hyperparameter} & \textbf{Concrete content} \\
		\midrule
		Model dimension & 256 \\
		Training epochs & 50 \\
		Number of heads & 4 \\
		Number of encoder layers & 2 \\
		Loss function & MSE \\
		Feed-forward network dimension & 512 \\
		Lookback window & 108 \\
		Initial learning rate & 0.0001 \\
		Batch size & 128 \\
		Optimizer & Adam \\
		Learning rate scheduler & Cosine Annealing LR \\
		Weight decay coefficient & 0.00001 \\
		Gradient clipping threshold & 0.5 \\
		\bottomrule
	\end{tabular}
\end{table}

To ensure training stability and prevent gradient explosion, we apply global gradient clipping \citep{pascanu2013difficulty} during the backpropagation of the feed-forward network. After computing the gradients $\hat{g}$ for all parameters in each backward pass, we constrain the L2 norm of the entire parameter gradient to remain below a predefined gradient clipping threshold $\tau$:
\begin{equation}
	\text{if } \|\hat{g}\|_{2} \geq \tau, \text{ then } \hat{g} \leftarrow \frac{\tau}{\|\hat{g}\|_{2}} \hat{g}.
\end{equation}
Simultaneously, to suppress model overfitting, we incorporate a weight decay strategy \citep{hanson1988comparing} into the optimizer. This strategy adds an L2 regularization term to the loss function:
\begin{equation}
	L_{\text{total}}(\theta) = \text{MSE}(\theta) + \frac{\lambda}{2} \|\theta\|_{2}^{2}.
\end{equation}
Where, $\|\theta\|_{2}^{2}$ represents the squared L2-norm of all model parameters $\theta$, $L_{\text{total}}(\theta)$ represents the total loss function after incorporating weight decay, $\text{MSE}(\theta)$ is the original loss function of mean squared error, and $\lambda$ is the weight decay coefficient that controls the regularization strength.

To objectively evaluate the prediction performance of the proposed method, we employ three evaluation indicators: Root Mean Square Error (RMSE), Mean Absolute Error (MAE), and Mean Absolute Percentage Error (MAPE). These indicators are among the most commonly used for time series forecasting. Their specific formulas are provided in Table \ref{tab2}. In our evaluation, the testing set contains $T$ distinct starting time points for prediction. For each starting time point $t$ ($t = 1,2,\ldots,T$), the model generates a forecast sequence for the next $S$ days. Here, $\hat{Y}_{t,s}$ signifies the predicted value for the $s$-th ($s = 1,2,\ldots,S$) day in the $t$-th ($t = 1,2,\ldots,T$) time point, and $Y_{t,s}$ represents the corresponding ground-truth value.

\begin{table}[H]
	\centering
	\caption{Formulas and descriptions of evaluation indicators.}
	\label{tab2}
	\renewcommand{\arraystretch}{1.5}
	\begin{tabular}{ccc}
		\toprule
		\textbf{Evaluation indicators} & \textbf{Formulas} & \textbf{Descriptions} \\
		\midrule
		RMSE & $\text{RMSE} = \sqrt{\frac{1}{T \times S} \sum_{t=1}^{T} \sum_{s=1}^{S} (\hat{Y}_{t,s} - Y_{t,s})^2}$ & Better closer to zero \\
		MAE & $\text{MAE} = \frac{1}{T \times S} \sum_{t=1}^{T} \sum_{s=1}^{S} |\hat{Y}_{t,s} - Y_{t,s}|$ & Better closer to zero \\
		MAPE & $\text{MAPE} = \frac{1}{T \times S} \sum_{t=1}^{T} \sum_{s=1}^{S} \left| \frac{\hat{Y}_{t,s} - Y_{t,s}}{Y_{t,s}} \right| \times 100\%$ & Better closer to zero \\
		\bottomrule
	\end{tabular}
\end{table}

\section{Results}\label{sec:Results} 
\subsection{Wavelet-decomposed data}\label{subsec:result1}
\begin{figure}[H]
	\centering
	\includegraphics[width=0.82\linewidth]{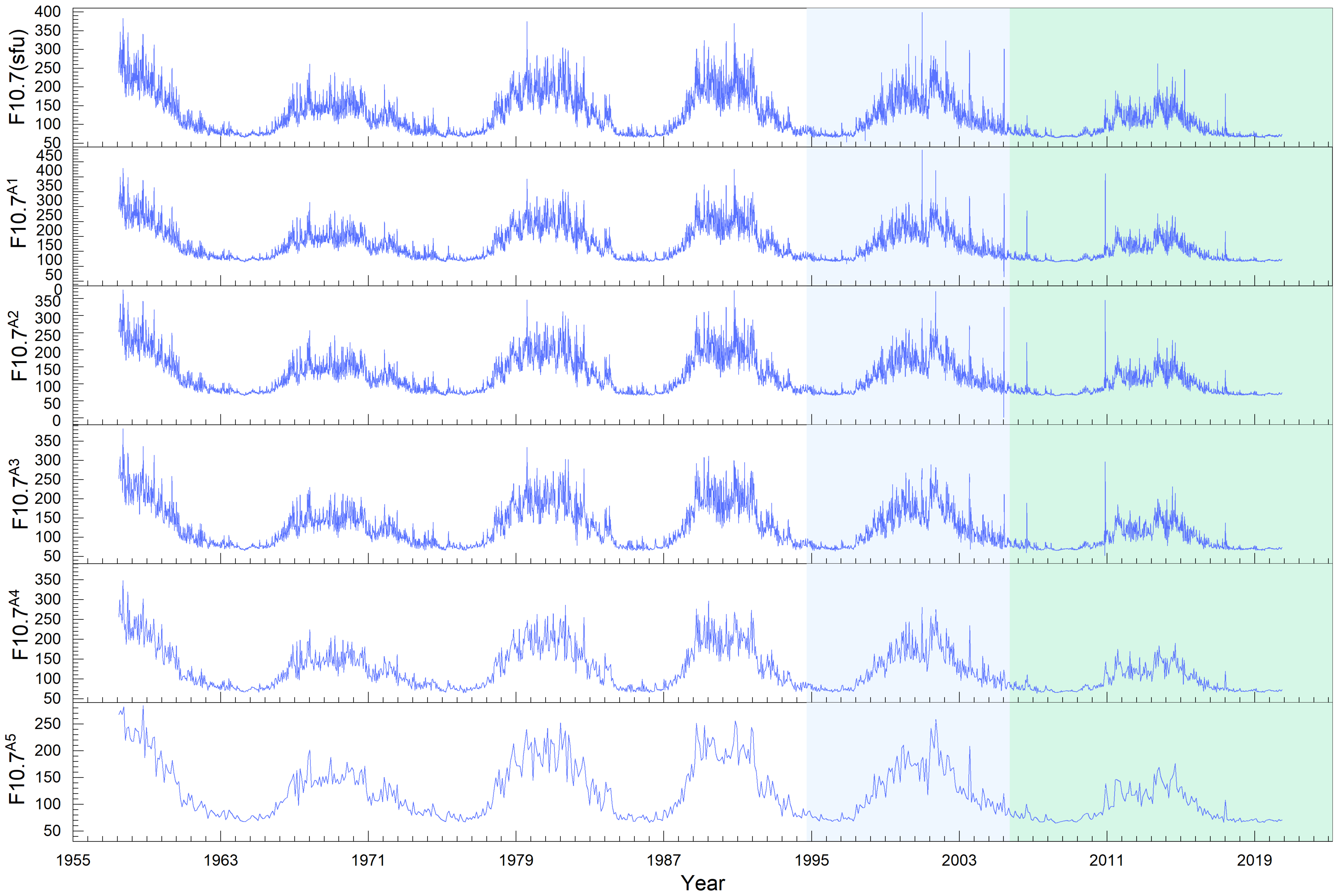}
	\caption{The wavelet-decomposed approximate signals of F10.7 index.}
	\label{fig6}
\end{figure}

\begin{figure}[H]
	\centering
	\includegraphics[width=0.82\linewidth]{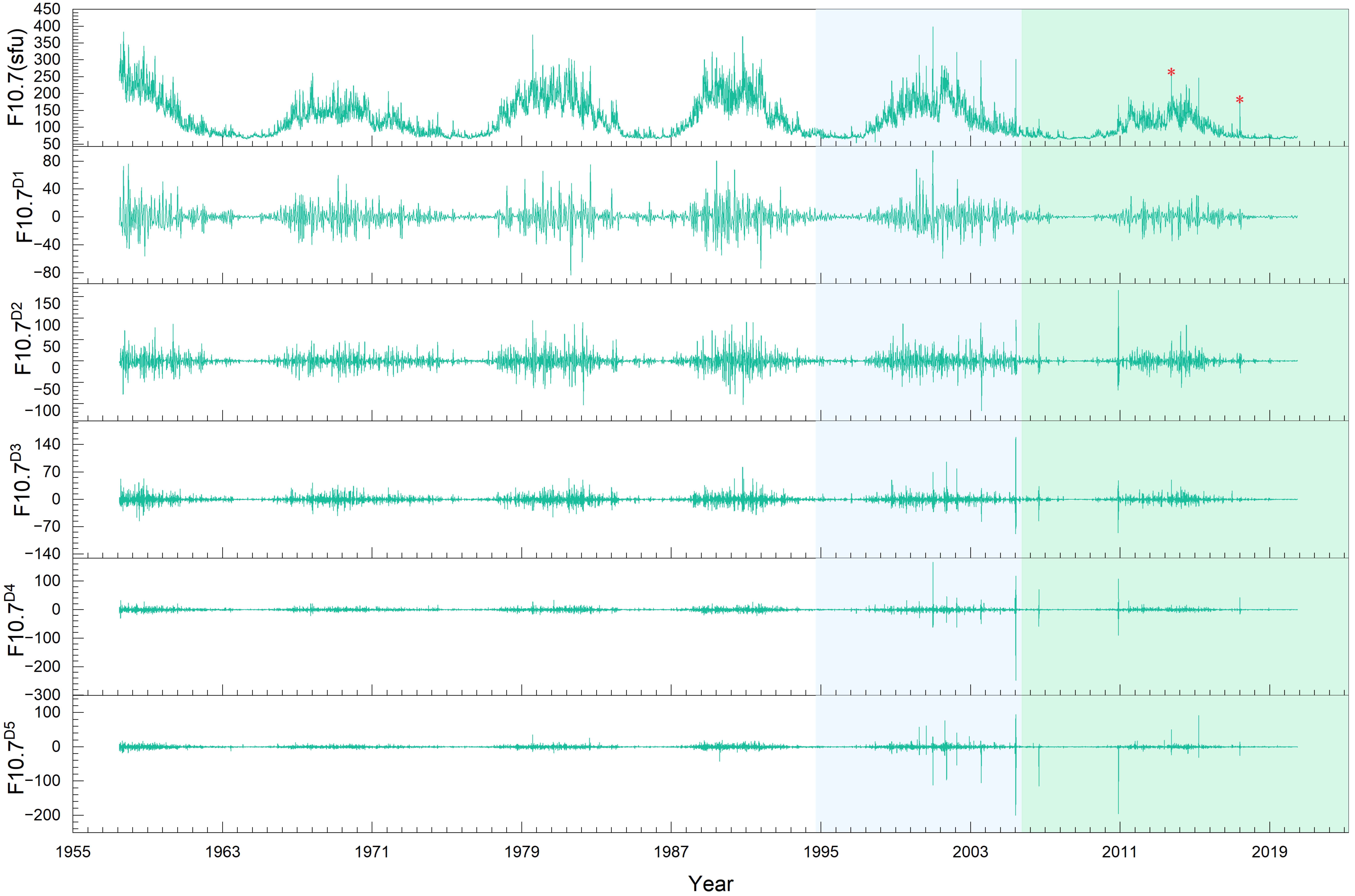}
	\caption{The wavelet-decomposed detail signals of F10.7 index.}
	\label{fig7}
\end{figure}

\begin{figure}[H]
	\centering
	\includegraphics[width=0.82\linewidth]{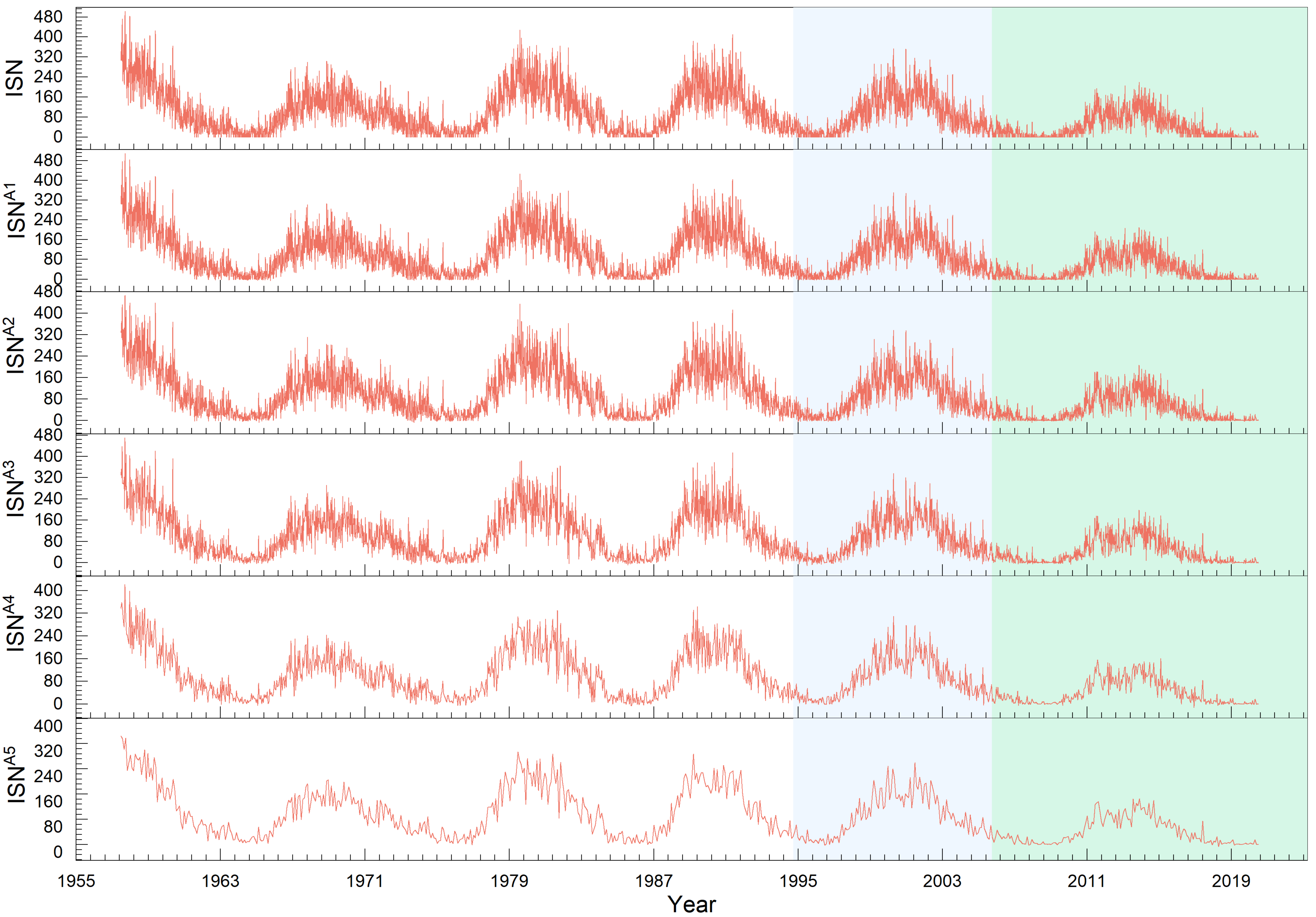}
	\caption{The wavelet-decomposed approximate signals of ISN.}
	\label{fig8}
\end{figure}

\begin{figure}[H]
	\centering
	\includegraphics[width=0.82\linewidth]{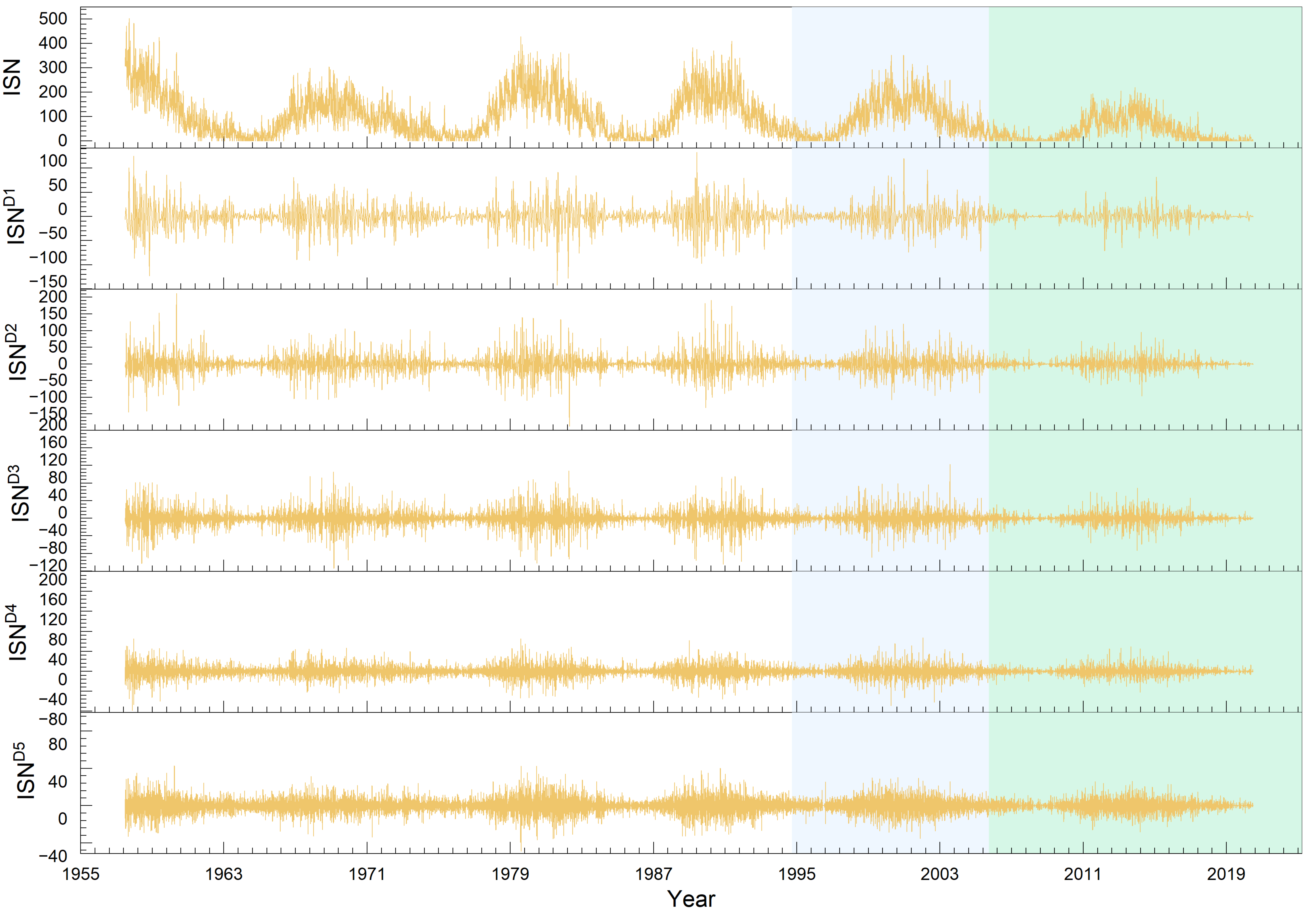}
	\caption{The wavelet-decomposed detail signals of ISN.}
	\label{fig9}
\end{figure}

We process the F10.7 index and ISN using wavelet decomposition. The decomposition level is the same as that of \citet{zucatelli2020nowcasting}. We then generate approximate signals and detail signals from the first to fifth level respectively, thereby constructing Dataset A. Ultimately, Dataset A comprises 22 feature variables: the F10.7 index, its first to fifth level approximate and detail signals, the ISN, and its corresponding first to fifth level approximate and detail signals. Figures \ref{fig6} and \ref{fig7} present the approximate and detail signals from the wavelet decomposition of the F10.7 index in the Dataset A, while Figures \ref{fig8} and \ref{fig9} show those for the ISN. The data with white, blue and green backgrounds in the figure correspond to the training, validation, and testing datasets, respectively.

Figures \ref{fig6} and \ref{fig8} show that as the decomposition level increases, the approximate signals of both the F10.7 index and the ISN become progressively smoother. This process helps to remove high-frequency noise and short-term fluctuations, thereby revealing the long-term trend of solar activity and highlighting the periodic patterns. Figures \ref{fig7} and \ref{fig9} show that the detail signals of the F10.7 index and the ISN generally exhibit low-amplitude variations within a narrow numerical range. However, these signals display distinct mutations or sharp peaks, which likely correspond to burst events in solar activity, such as solar flares, coronal mass ejections, or other space weather phenomena. According to observational data from the U.S. Department of Commerce, NOAA, Space Weather Prediction Center (\url{ftp://ftp.swpc.noaa.gov/pub/warehouse/}), the two dates marked with red star symbols in Figure \ref{fig7}, January 7, 2014, and September 4, 2017, both experienced intense solar activity. On January 7, 2014, solar activity was intense, with detail signals peaking. The F10.7 index for that day reached as high as 237.1 sfu. Active region (AR) 11944, AR11946, and AR11947 collectively produced 1 X-class, 2 M-class, and 10 C-class flares. Similarly, on September 4, 2017, solar activity was frequent, and the detail signals showed abrupt changes, with a corresponding F10.7 index of 182.5 sfu. There was a total of 7 M-class flares, 9 C-class flares and 2 CMEs on that day.

\subsection{Performance comparison: integrating wavelet-decomposed F10.7 signals}\label{subsec:result2}
In this section, we investigate the impact of different combinations of wavelet-decomposed signals on model performance. Table \ref{tab3} presents different combinations that integrates the approximate signals and detail signals of the F10.7 index. For each combination, we train, validate, and test the model on Dataset A. Table \ref{tab4} shows the performance results for each combination method on the corresponding testing datasets under different prediction lengths, with bold values indicating the best performance for each prediction length. Figure \ref{fig10} provides a visual representation of the data from Table \ref{tab4}.

\begin{table}[htbp]
	\centering
	\caption{Different combinations that integrates the approximate signals and detail signals of the F10.7 index.}
	\label{tab3}
	\begin{tabular}{ccc}
		\toprule
		Combination & Number of & The content of the combination \\
		number & variables & \\
		\midrule
		1 & 1 & Original F10.7 index \\
		2 & 3 & \begin{tabular}[c]{@{}c@{}}The original F10.7 index with its first level approximate signals and detail signals\\
			(F10.7, F10.7$^\mathrm{A1}$, F10.7$^\mathrm{D1}$)\end{tabular} \\
		3 & 5 & \begin{tabular}[c]{@{}c@{}}The original F10.7 index with its first to second level approximate signals and detail signals \\
			(F10.7, F10.7$^\mathrm{A1}$, F10.7$^\mathrm{D1}$, F10.7$^\mathrm{A2}$, F10.7$^\mathrm{D2}$)\end{tabular} \\
		4 & 7 & \begin{tabular}[c]{@{}c@{}}The original F10.7 index with its first to third level approximate signals and detail signals \\
			(F10.7, F10.7$^\mathrm{A1}$, F10.7$^\mathrm{D1}$, $\cdots$, F10.7$^\mathrm{A3}$, F10.7$^\mathrm{D3}$)\end{tabular} \\
		5 & 9 & \begin{tabular}[c]{@{}c@{}}The original F10.7 index with its first to fourth level approximate signals and detail signals \\
			(F10.7, F10.7$^\mathrm{A1}$, F10.7$^\mathrm{D1}$, $\cdots$, F10.7$^\mathrm{A4}$, F10.7$^\mathrm{D4}$)\end{tabular} \\
		6 & 11 & \begin{tabular}[c]{@{}c@{}}The original F10.7 index with its first to fifth level approximate signals and detail signals \\
			(F10.7, F10.7$^\mathrm{A1}$, F10.7$^\mathrm{D1}$, $\cdots$, F10.7$^\mathrm{A5}$, F10.7$^\mathrm{D5}$)\end{tabular} \\
		\bottomrule
	\end{tabular}
\end{table}

\begin{table}[H]
	\centering
	\caption{Performance results for each combination method on the corresponding testing datasets under different prediction lengths.}
	\label{tab4}
	\setlength{\tabcolsep}{1pt} 
	\small
	\begin{tabular}{c *{14}{c}}
		\toprule
		\multirow{2}{*}{Evaluation indicators} & \multirow{2}{*}{Combination} & \multicolumn{12}{c}{Prediction length (days)} \\
		\cmidrule(lr){3-15}
		& & 1 & 2 & 3 & 5 & 10 & 15 & 20 & 25 & 27 & 30 & 35 & 40 & 45 \\
		\midrule
		\multirow{6}{*}{RMSE}
		& Combination 1 & 5.236 & 6.089 & 6.919 & 8.397 & 11.875 & 11.768 & 12.349 & 13.633 & 13.143 & 13.298 & 13.775 & 14.320 & 14.495 \\
		& Combination 2 & 4.547 & 5.416 & 6.618 & 8.758 & 11.189 & 12.063 & 13.197 & 13.298 & 13.639 & 13.947 & 13.307 & 14.550 & 13.868 \\
		& Combination 3 & 4.204 & 4.695 & 5.700 & 7.348 & 10.469 & 11.565 & 12.987 & 11.663 & 14.076 & 12.004 & 13.497 & 12.666 & 13.708 \\
		& Combination 4 & \textbf{4.165} & \textbf{4.445} & 5.290 & \textbf{5.604} & 8.083 & 10.090 & 10.889 & 11.305 & 12.519 & 12.689 & 12.442 & 13.781 & 14.573 \\
		& Combination 5 & 4.232 & 4.484 & 5.189 & 6.027 & \textbf{7.974} & 10.056 & \textbf{10.352} & 11.606 & 11.975 & 12.421 & \textbf{12.438} & 13.126 & 13.658 \\
		& Combination 6 & 4.526 & 4.587 & \textbf{5.173} & 6.125 & 8.039 & \textbf{9.614} & 10.571 & \textbf{10.149} & \textbf{11.145} & \textbf{11.499} & 12.457 & \textbf{12.580} & \textbf{12.511} \\
		\midrule
		\multirow{6}{*}{MAE}
		& Combination 1 & 2.541 & 3.148 & 3.643 & 4.591 & 6.866 & 6.896 & 7.371 & 8.225 & 7.919 & 8.017 & 8.387 & 8.791 & 8.884 \\
		& Combination 2 & 2.295 & 2.895 & 3.697 & 4.901 & 6.413 & 6.957 & 7.790 & 7.956 & 8.141 & 8.280 & 8.055 & 8.739 & 8.382 \\
		& Combination 3 & \textbf{2.091} & 2.484 & 3.029 & 4.000 & 5.921 & 6.760 & 7.646 & 6.845 & 8.243 & 7.109 & 8.066 & 7.548 & 8.270 \\
		& Combination 4 & 2.116 & 2.343 & 2.875 & \textbf{3.088} & 4.632 & 5.804 & 6.281 & 6.536 & 7.276 & 7.469 & 7.352 & 8.225 & 8.848 \\
		& Combination 5 & 2.228 & \textbf{2.314} & 2.820 & 3.248 & \textbf{4.429} & 5.557 & \textbf{5.922} & 6.735 & 6.947 & 7.260 & 7.341 & 7.769 & 8.255 \\
		& Combination 6 & 2.225 & 2.381 & \textbf{2.644} & 3.327 & 4.562 & \textbf{5.334} & 5.985 & \textbf{5.901} & \textbf{6.455} & \textbf{6.721} & \textbf{7.330} & \textbf{7.398} & \textbf{7.422} \\
		\midrule
		\multirow{6}{*}{MAPE}
		& Combination 1 & 2.345 & 2.915 & 3.373 & 4.240 & 6.362 & 6.392 & 6.865 & 7.650 & 7.360 & 7.442 & 7.804 & 8.223 & 8.272 \\
		& Combination 2 & 2.139 & 2.693 & 3.431 & 4.527 & 5.958 & 6.474 & 7.168 & 7.353 & 7.550 & 7.672 & 7.540 & 8.137 & 7.785 \\
		& Combination 3 & \textbf{1.950} & 2.317 & 2.816 & 3.706 & 5.500 & 6.266 & 7.055 & 6.316 & 7.594 & 6.643 & 7.502 & 6.995 & 7.704 \\
		& Combination 4 & 1.975 & 2.197 & 2.682 & \textbf{2.883} & 4.318 & 5.380 & 5.818 & 6.045 & 6.719 & 6.932 & 6.834 & 7.646 & 8.282 \\
		& Combination 5 & 2.084 & \textbf{2.165} & 2.627 & 2.993 & \textbf{4.083} & 5.079 & \textbf{5.457} & 6.239 & 6.425 & 6.711 & 6.812 & 7.197 & 7.767 \\
		& Combination 6 & 2.062 & 2.223 & \textbf{2.458} & 3.100 & 4.220 & \textbf{4.883} & 5.486 & \textbf{5.433} & \textbf{5.954} & \textbf{6.200} & \textbf{6.779} & \textbf{6.859} & \textbf{6.891} \\
		\bottomrule
	\end{tabular}
\end{table}

As shown in Table \ref{tab4}, we compare the prediction performance of different combination methods under various prediction lengths on Dataset A. Taking Combination 1, which uses only the original F10.7 index, as the baseline method, we compare its performance with that of other combination methods. Using the Combination 6 method as an example, the performance with a prediction length of 45 days shows significant improvement compared to the baseline method. Specifically, RMSE decreases by 13.69\%, MAE by 16.45\%, and MAPE by 16.69\%. As illustrated in Figure \ref{fig10}, compared to the baseline method using only the original F10.7 index, the combination methods integrating wavelet-decomposed signals demonstrate overall improvement in prediction performance. When the prediction length is 1, 2, 5 and 10 days, the prediction performance of the Combination 6 method is slightly inferior to that of the Combination 3, 4 and 5 methods. However, as the prediction length increases, the advantage of the Combination 6 method, which includes higher-level wavelet-decomposed signals, becomes increasingly pronounced. As higher-level approximate and detail signals are added, the prediction performance shows an overall upward trend. The prediction performance for all combination methods decreases as the prediction length increases. Overall, compared with other combination methods, the Combination 6 method achieves outstanding prediction performance. We speculate that this might be because Combination 6 contains more abundant multi-scale feature information. The approximate signals correspond to the low-frequency changes of the F10.7 index, while the detail signals correspond to the high-frequency fluctuations, allowing the model to simultaneously capture the features at different frequencies. After a comprehensive comparison of the Combinations 2 to 6, we recommend using the Combination 6 method, which includes first to fifth level signals.

\begin{figure}[H]
	\centering
	\includegraphics[width=0.79\linewidth]{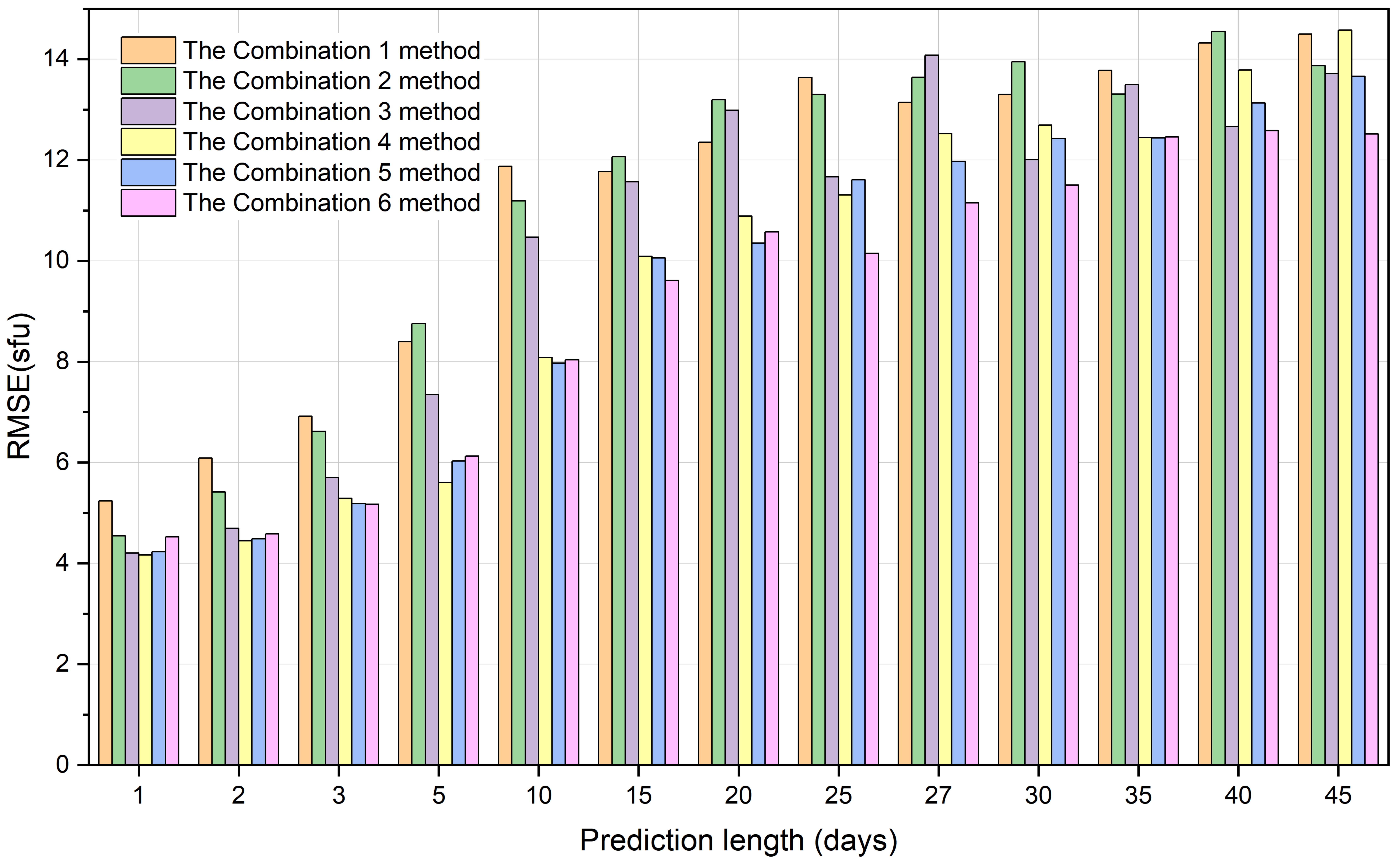}
\end{figure}
\begin{figure}[H]
	\centering
	\includegraphics[width=0.79\linewidth]{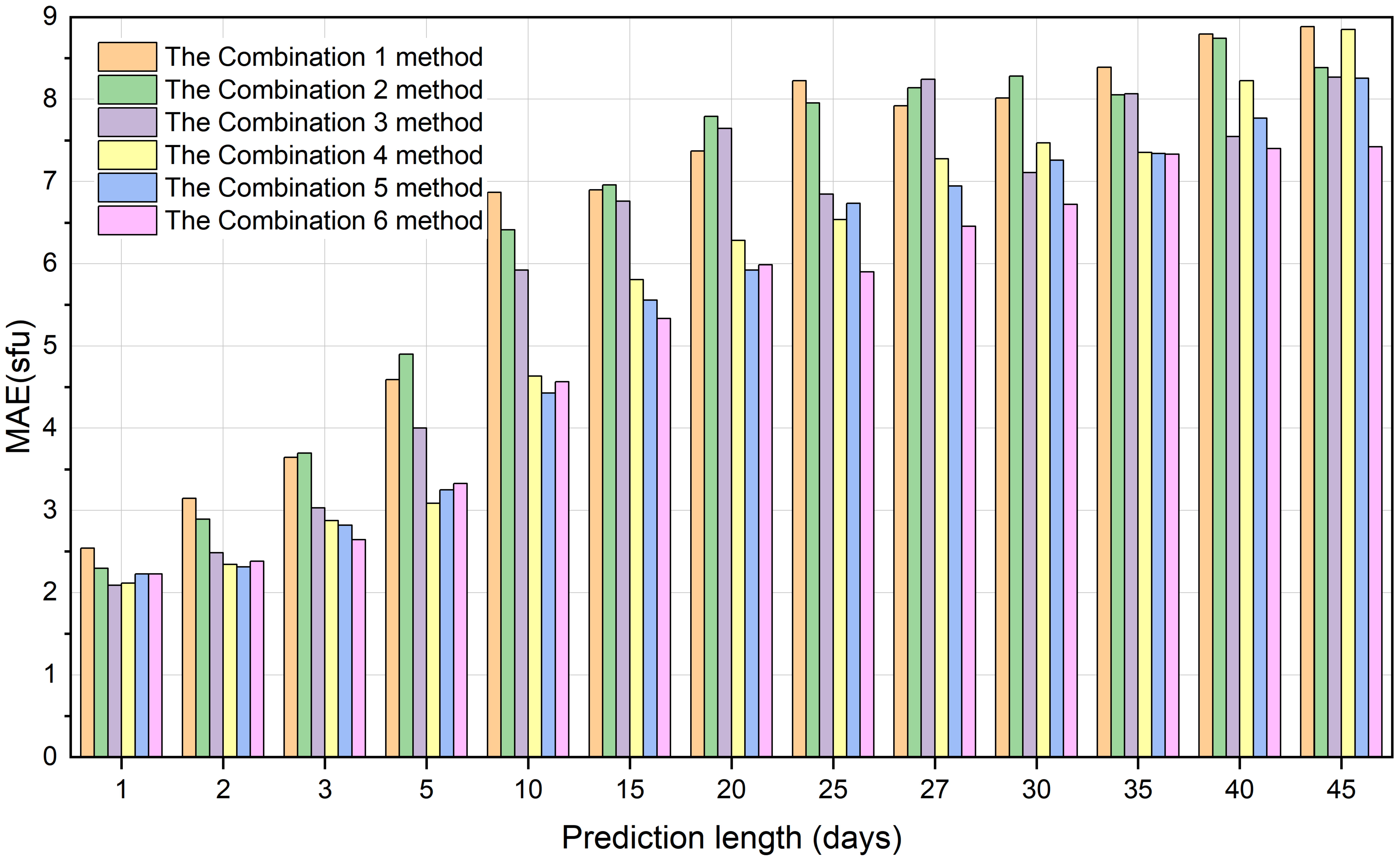}
\end{figure}
\begin{figure}[H]
	\centering
	\includegraphics[width=0.79\linewidth]{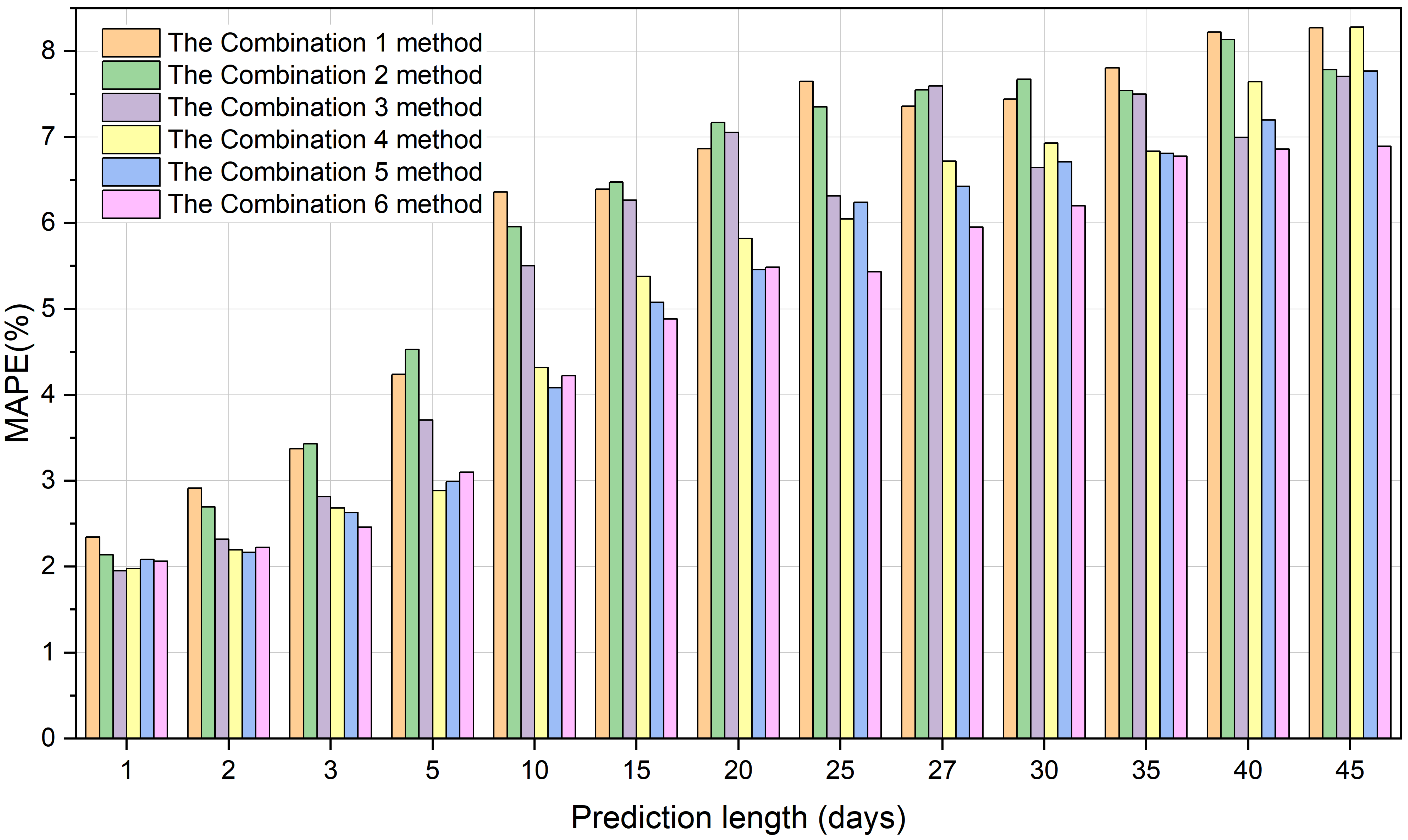}
	\caption{Bar chart comparing the prediction performance results for Combinations 1 to 6 methods under different prediction lengths.}
	\label{fig10}
\end{figure}

\begin{figure}[htbp]
	\centering
	\includegraphics[width=0.45\linewidth]{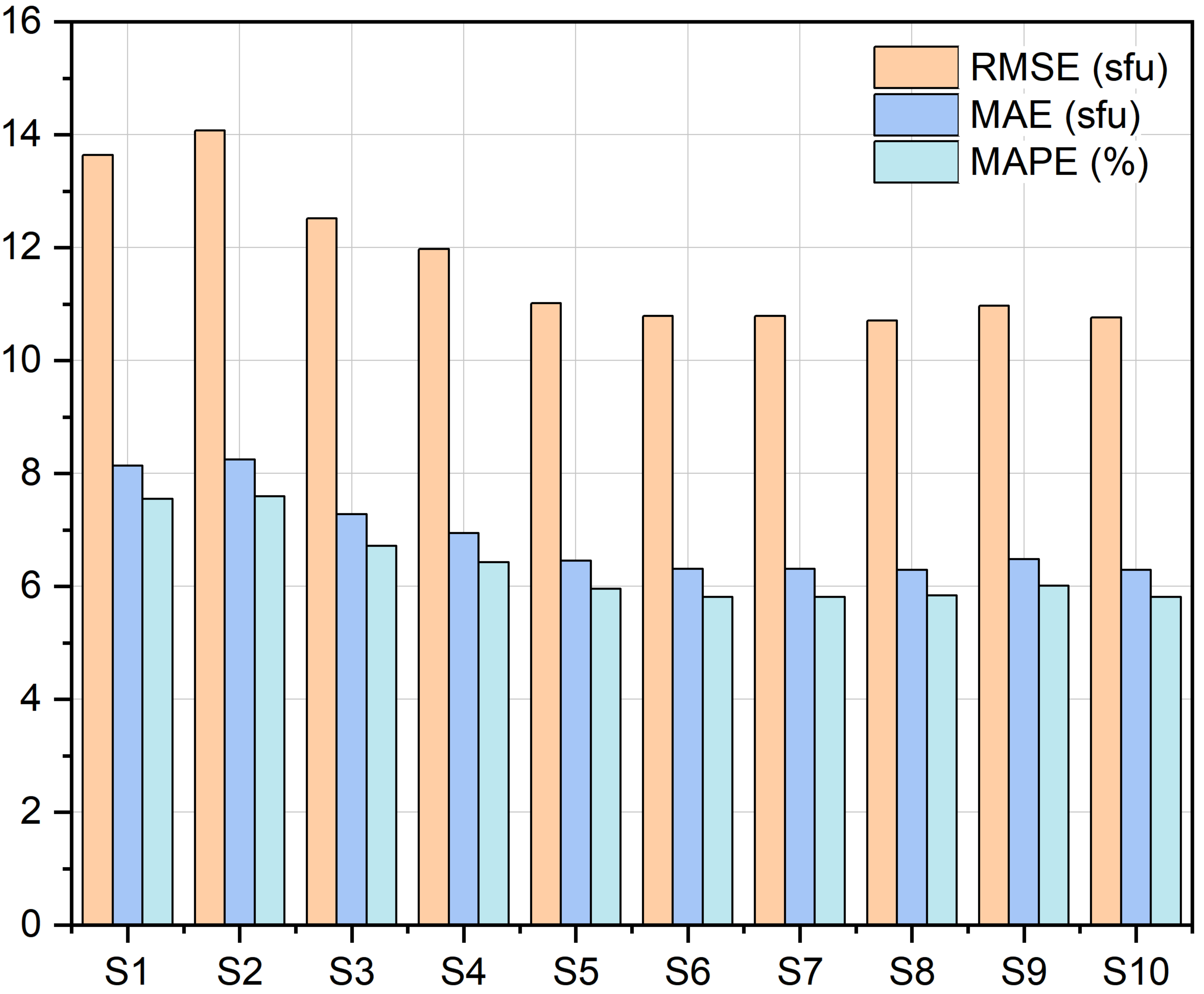}
	\caption{Bar chart comparing the prediction performance of methods that incorporate different level of wavelet-decomposed signals when the prediction length is 27 days.}
	\label{fig10+}
\end{figure}

As higher-level approximate signals and detail signals are added, the overall performance shows an improving trend. To verify whether the Combination 6 is the optimal combination, we sequentially accumulate approximate and detail signals of the sixth, seventh, eighth, ninth, and tenth level based on the Combination 6 and then investigate their impact on the prediction performance of model. Figure \ref{fig10+} presents bar chart comparing the prediction performance of methods that incorporate different level of wavelet-decomposed signals when the prediction length is 27 days. In the figure, S1 corresponds to the Combination 2, S2 corresponds to the Combination 3, and so on, with S5 corresponding to the Combination 6. S6 to S10 sequentially add approximate and detail signals of the sixth to tenth level on the basis of S5. For example, S10 includes the original F10.7 index and its first to tenth level approximate and detail signals, totaling 21 variables. As shown in Figure \ref{fig10+}, among S1 to S5, the S5 method achieves the best prediction performance, which is consistent with the conclusion drawn earlier. However, from S5 to S10, the prediction performance of all methods show no significant improvement. This result indicates that introducing wavelet-decomposed signals of the first to fifth levels can effectively enhance prediction performance of model, but further adding signals of sixth level and above does not contribute to gains in prediction performance. Therefore, Combination 6, which includes first to fifth level signals, is the optimal combination.

After identifying the Combination 6 method, which incorporates the original F10.7 index combined with its first to fifth level approximate signals and detail signals, as the optimal method, we further investigate the impact of the methods for approximate signal combination, detail signal combination, and Combination 6 on model performance. Specifically, we separately construct two combinations. The first is the approximate signal combination (Combination A), which includes the original F10.7 index and its first to fifth level approximate signals. The second is the detail signal combination (Combination D), which includes the original F10.7 index and its first to fifth level detail signals. We then compare their prediction performance with that of the Combination 6 method. Table \ref{tab5} presents the performance results for the methods of Combination 6, Combination A and Combination D on the corresponding testing datasets under different prediction lengths, with bold values indicating the best performance for each prediction length. Figure \ref{fig11} provides a visual representation of the data from Table \ref{tab5}.

\begin{figure}[H]
	\centering
	\includegraphics[width=0.5\linewidth]{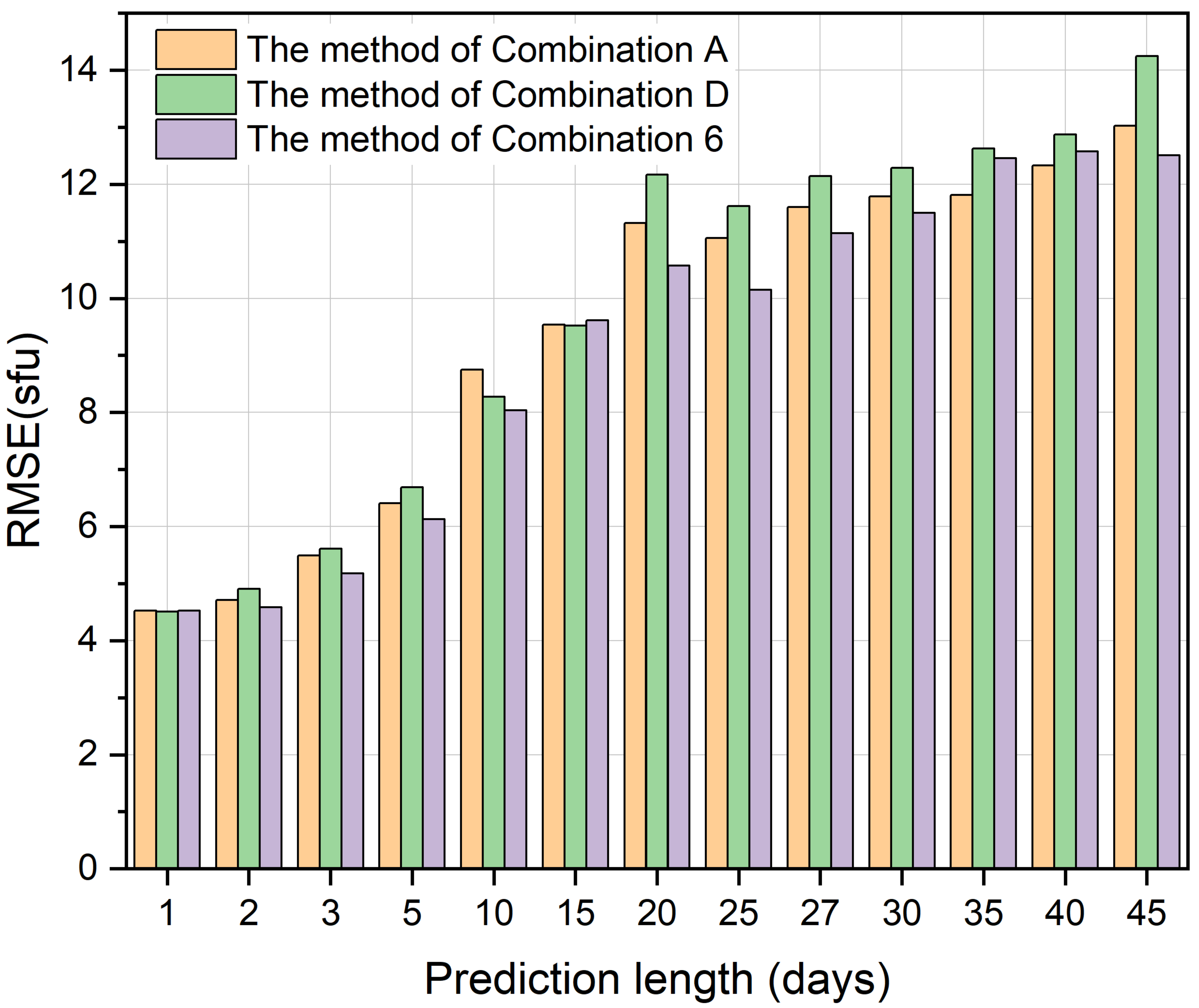}
\end{figure}
\begin{figure}[H]
	\centering
	\includegraphics[width=0.5\linewidth]{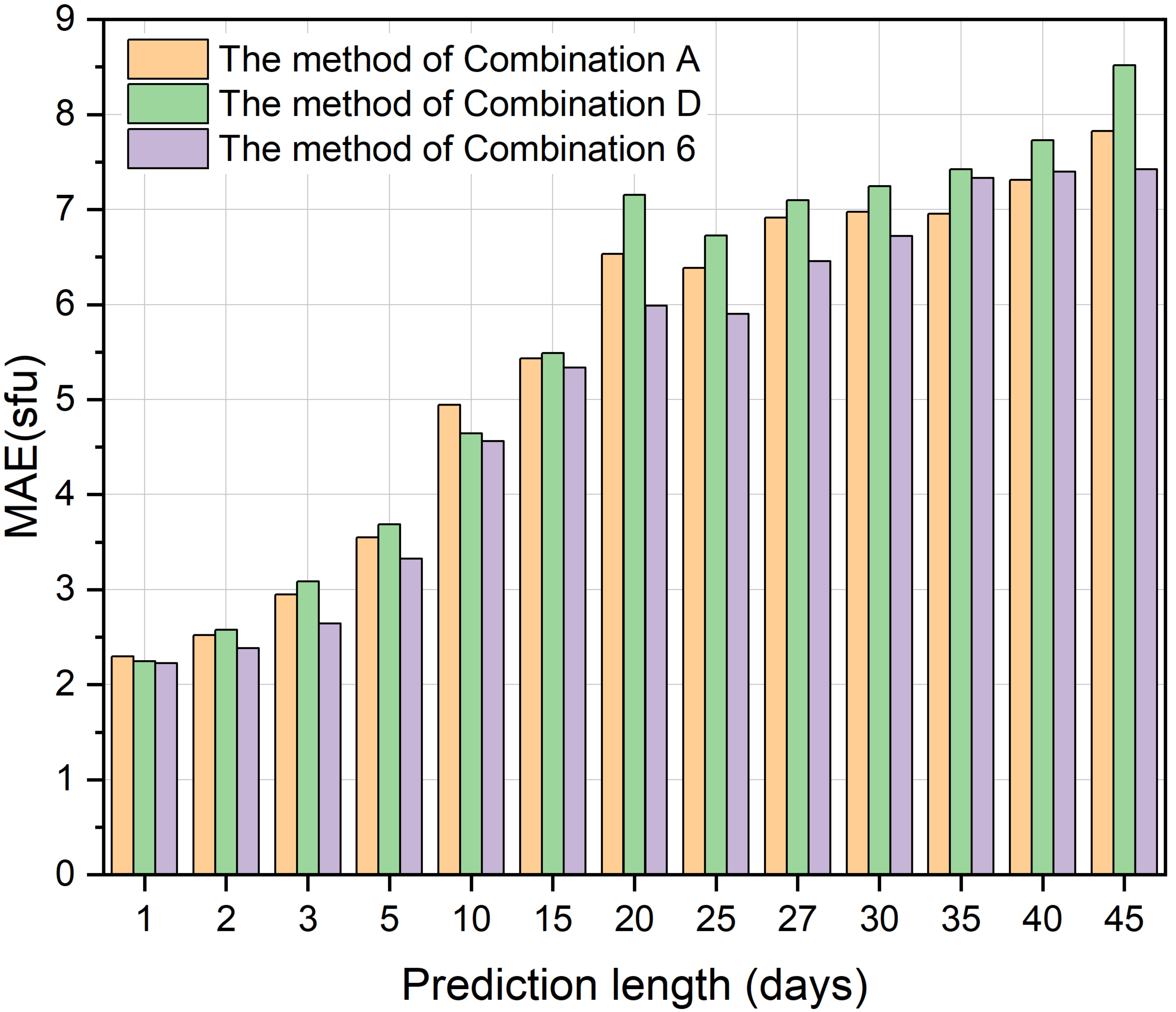}
\end{figure}
\begin{figure}[H]
	\centering
	\includegraphics[width=0.5\linewidth]{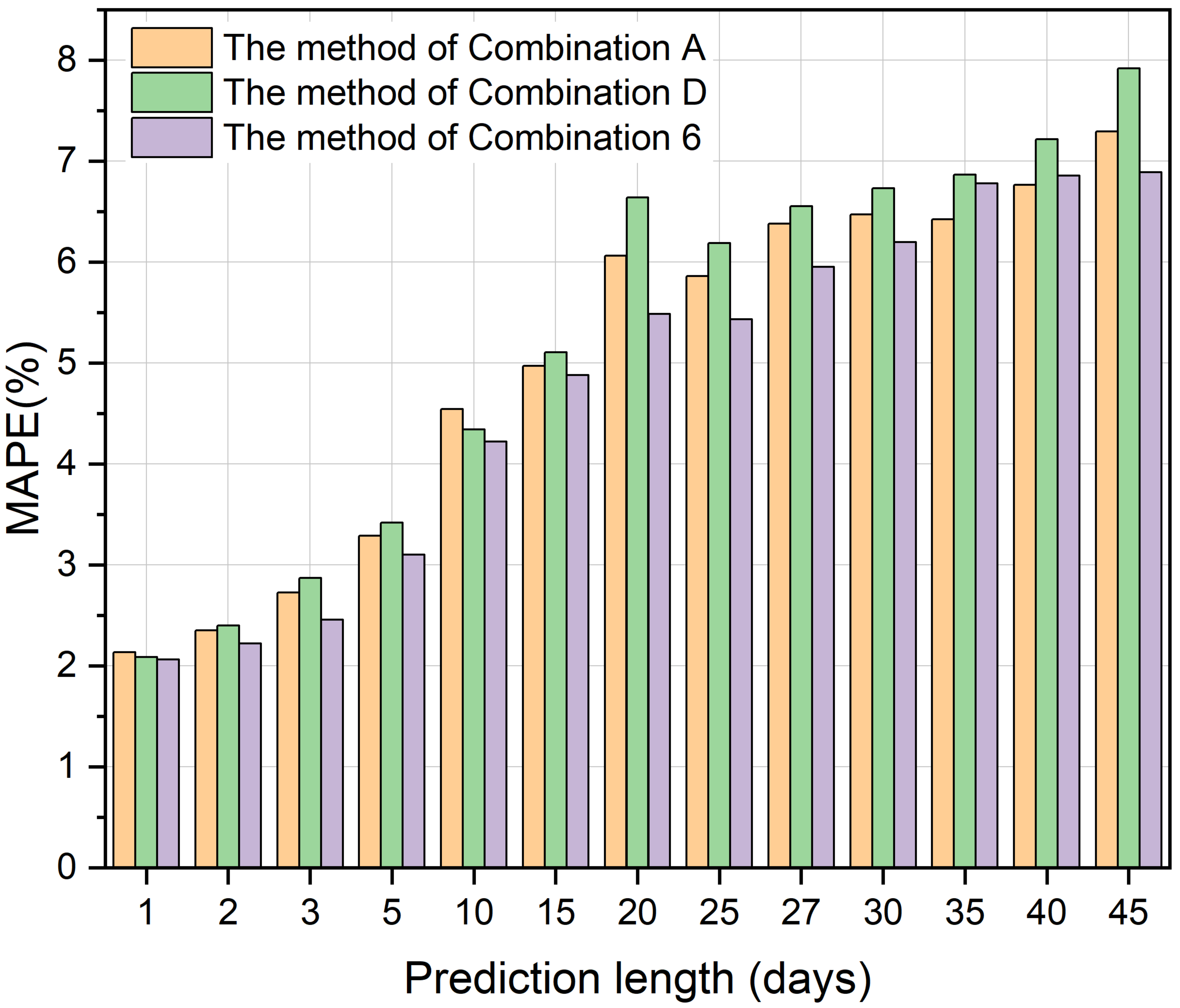}
	\caption{Bar chart comparing the prediction performance results for the methods of Combination 6, Combination A and Combination D under different prediction lengths.}
	\label{fig11}
\end{figure}

\begin{table}[h]
	\centering
	\caption{Performance results for the methods of Combination 6, Combination A and Combination D on the corresponding testing datasets under different prediction lengths.}
	\label{tab5}
	\setlength{\tabcolsep}{4pt}
	\small
	\begin{tabular}{c *{9}{c}}
		\toprule
		\multirow{2}{*}{Prediction length (days)} & \multicolumn{3}{c}{Combination 6} & \multicolumn{3}{c}{Combination A} & \multicolumn{3}{c}{Combination D} \\
		\cmidrule(lr){2-10}
		& RMSE & MAE & MAPE & RMSE & MAE & MAPE & RMSE & MAE & MAPE \\
		\midrule
		1  & 4.526&	\textbf{2.225}&\textbf{2.062}&	4.525&	2.295&	2.134&	\textbf{4.505}&	2.246&	2.089 \\
		2  & \textbf{4.587}&	\textbf{2.381}&	\textbf{2.223}&	4.711	&2.523&2.353&	4.908	&2.577&2.401 \\
		3  & \textbf{5.173}&\textbf{2.644}&	\textbf{2.458}&	5.487	&2.949	&2.727	&5.608&	3.088&	2.871 \\
		5  & \textbf{6.125}	&\textbf{3.327}&	\textbf{3.100}&	6.404	&3.552	&3.290	&6.691	&3.688	&3.420 \\
		10 & \textbf{8.039}	&\textbf{4.562}&	\textbf{4.220}&	8.753&	4.946&	4.545&	8.276	&4.645	&4.343 \\
		15 & 9.614&	\textbf{5.334}	&\textbf{4.883}	&9.538&	5.430&	4.971	&\textbf{9.522}&	5.488	&5.107 \\
		20 & \textbf{10.571}	&\textbf{5.985}	&\textbf{5.486}&	11.324&	6.531	&6.062	&12.165	&7.154	&6.643 \\
		25 & \textbf{10.149}&	\textbf{5.901}&	\textbf{5.433}	&11.060	&6.383	&5.864	&11.613	&6.724	&6.189 \\
		27 & \textbf{11.145}&	\textbf{6.455}&	\textbf{5.954}	&11.597	&6.912	&6.381	&12.142	&7.099	&6.553 \\
		30 & \textbf{11.499}	&\textbf{6.721}&\textbf{6.200}	&11.782	&6.977&	6.470&	12.289	&7.245	&6.731 \\
		35 & \textbf{12.457}	&\textbf{7.330}	&\textbf{6.779}	&11.808	&6.956&	6.426	&12.627	&7.422	&6.868 \\
		40 & \textbf{12.580}&	\textbf{7.398}&	\textbf{6.859}	&12.325	&7.309	&6.767	&12.877	&7.726	&7.218 \\
		45 & \textbf{12.511}	&\textbf{7.422}	&\textbf{6.891}&	13.028	&7.824&	7.293	&14.243	&8.517&	7.920 \\
		\bottomrule
	\end{tabular}
\end{table}

As shown in Table \ref{tab5}, we compare the prediction performance results for the methods of Combination 6, Combination A, and Combination D under different prediction lengths. Taking the prediction length of 45 days as an example, compared with the Combination A method, the RMSE for the Combination 6 method decreases by 12.75\%, the MAE decreases by 15.11\%, and the MAPE decreases by 15.95\%. Compared with the Combination D method, the RMSE for the Combination 6 method decreases by 15.04\%, the MAE decreases by 14.23\%, and the MAPE decreases by 11.76\%. Compared with the Combination D method, the RMSE for the Combination A method decreases by 8.23\%, the MAE decreases by 8.67\%, and the MAPE decreases by 8.79\%. As shown in Figure \ref{fig11}, the prediction performance of the Combination 6 method is superior to that of the Combination D method and Combination A method. The prediction performance of the Combination A method is superior to that of the Combination D method. In summary, the Combination 6 method exhibits the optimal prediction performance, and the prediction performance of the Combination A method is superior to that of the Combination D. This is mainly because the approximate signals contribute more than the detail signals. The change of the F10.7 index is mainly dominated by the long-term trend of solar activity. Although the detail signals contain information such as solar flares caused by instantaneous disturbances, their complexity and uncertainty are relatively high. Therefore, the prediction performance of the Combination A method is superior to that of the Combination D method. The Combination 6 method can achieve the optimal performance, revealing the synergistic effect of the approximate signals that reflect the long-term trend of solar activity and the detail signals that represent short-term suddenness in the prediction process. Although the approximate signals provide the main prediction information, the introduction of the detail signals plays a key fine-tuning role, which may enable the model to have the ability to precisely capture the sudden changes and short-term fluctuations in actual observations that deviate from the smooth trend. This synergistic effects of approximate signals and detail signals enable the model to not only grasp the overall pattern of solar activity but also utilize the detail features, thereby achieving an improvement in prediction performance.

\subsection{Performance comparison: integrating wavelet-decomposed ISN signals}\label{subsec:result3}
In this section, we further investigate the impact of incorporating ISN and its approximate and detail signals on prediction performance, building upon the Combination 6, which incorporates the original F10.7 index combined with its first to fifth level approximate signals and detail signals. For different combination methods, we train, validate, and test the model on Dataset A. Previous studies have demonstrated that ISN can be used for F10.7 index prediction \citep{okoh2020relationships, henney2012forecasting}, and there exists a close intrinsic relationship between ISN and the F10.7 index \citep{shen2025multiscale,bhargawa2021elucidation}. Based on this foundation, we aim to explore the impact of introducing ISN and its approximate and detail signals on the prediction performance of F10.7 index. Table \ref{tab6} presents different combinations that integrates the ISN along with its approximate signals and detail signals. Figure \ref{fig12} visualizes the performance results for the Combination 6 method and the Combination 7 to 12 methods on the corresponding testing datasets under different prediction lengths.

\setlength{\tabcolsep}{1pt}
\begin{table}[h]
	\centering
	\caption{Different combinations that integrates ISN along with its approximate signals and detail signals.}
	\label{tab6}
	\begin{tabular}{ccc}
		\toprule
		Combination & Number of & The content of the combination \\
		number & variables & \\
		\midrule
		7 & 12 & Combination 6, ISN \\
		8 & 14 & \begin{tabular}[c]{@{}c@{}}Combination 6, as well as the ISN and its first level approximate signals and detail signals\\
			(Combination 6, ISN, ISN$^\mathrm{A1}$, ISN$^\mathrm{D1}$)\end{tabular} \\
		9 & 16 & \begin{tabular}[c]{@{}c@{}}Combination 6, as well as the ISN and its first to second level approximate signals and detail signals \\
			(Combination 6, ISN, ISN$^\mathrm{A1}$, ISN$^\mathrm{D1}$, ISN$^\mathrm{A2}$, ISN$^\mathrm{D2}$)\end{tabular} \\
		10 & 18 & \begin{tabular}[c]{@{}c@{}}Combination 6, as well as the ISN and its first to third level approximate signals and detail signals \\
			(Combination 6, ISN, ISN$^\mathrm{A1}$, ISN$^\mathrm{D1}$, $\cdots$, ISN$^\mathrm{A3}$, ISN$^\mathrm{D3}$)\end{tabular} \\
		11 & 20 & \begin{tabular}[c]{@{}c@{}}Combination 6, as well as the ISN and its first to fourth level approximate signals and detail signals \\
			(Combination 6, ISN, ISN$^\mathrm{A1}$, ISN$^\mathrm{D1}$, $\cdots$, ISN$^\mathrm{A4}$, ISN$^\mathrm{D4}$)\end{tabular} \\
		12 & 22 & \begin{tabular}[c]{@{}c@{}}Combination 6, as well as the ISN and its first to fifth level approximate signals and detail signals \\
			(Combination 6, ISN, ISN$^\mathrm{A1}$, ISN$^\mathrm{D1}$, $\cdots$, ISN$^\mathrm{A5}$, ISN$^\mathrm{D5}$)\end{tabular} \\
		\bottomrule
	\end{tabular}
\end{table}

As shown in Figure \ref{fig12}, we compare the prediction performance of the Combination 6 method and the Combination 7 to 12 methods on Dataset A under different prediction lengths. The prediction performance of the Combination 7 to 12 methods is comparable to that of the Combination 6 method. However, previous studies have shown that introducing the ISN can enhance the prediction performance \citep{henney2012forecasting}. We calculate that the Spearman correlation coefficient between the F10.7 index and ISN, as used in our study, is as high as 0.954. Therefore, we speculate that the information contained in them might be overlapping, which leads to no improvement in the prediction performance. Further, the wavelet-decomposed signals of F10.7 index can provide rich multi-scale feature information, which may have fully characterized the variation patterns of the F10.7 index. Specifically, we decompose the F10.7 index into characteristic signals at different frequencies through wavelet decomposition. The approximate signals reflect the long-term slow-changing trend of the F10.7 index, while the detail signals correspond to high-frequency fluctuations. These signals can comprehensively represent the change patterns of the F10.7 index and provide more feature information for the model, so introducing ISN and its wavelet-decomposed signals into the Combined 6 method does not improve the prediction performance. Additionally, it is worth noting that the relationship between the F10.7 index and ISN varies with different solar activity phases \citep{okoh2020relationships}. Therefore, the method of adding ISN and its wavelet-decomposed signals to Combination 6 does not enhance prediction performance in this study. In subsequent research, we will only consider the Combination 6 method, which demonstrates the optimal prediction performance.

\begin{figure}[H]
	\centering
	\includegraphics[width=0.8\linewidth]{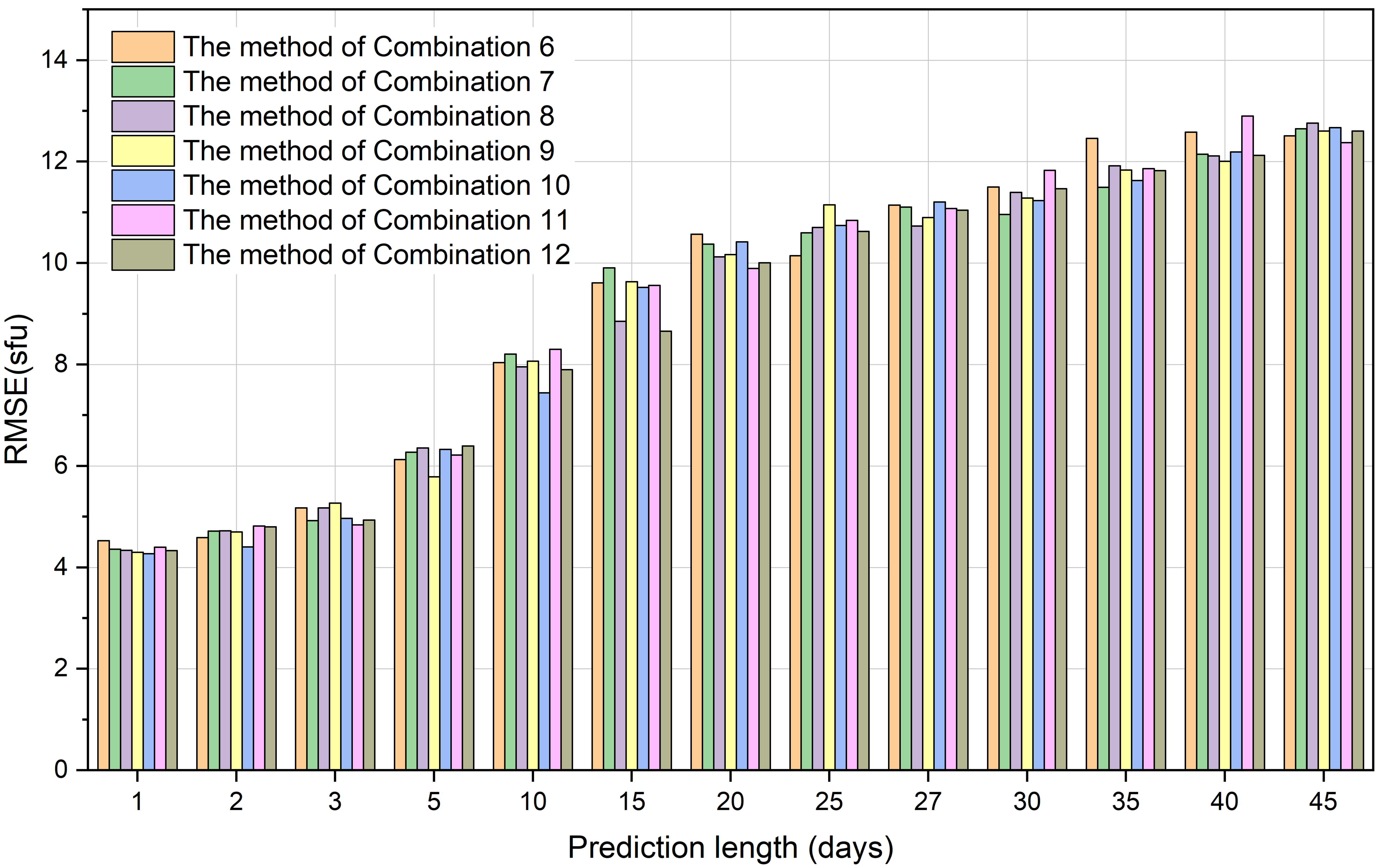}
\end{figure}
\begin{figure}[H]
	\centering
	\includegraphics[width=0.8\linewidth]{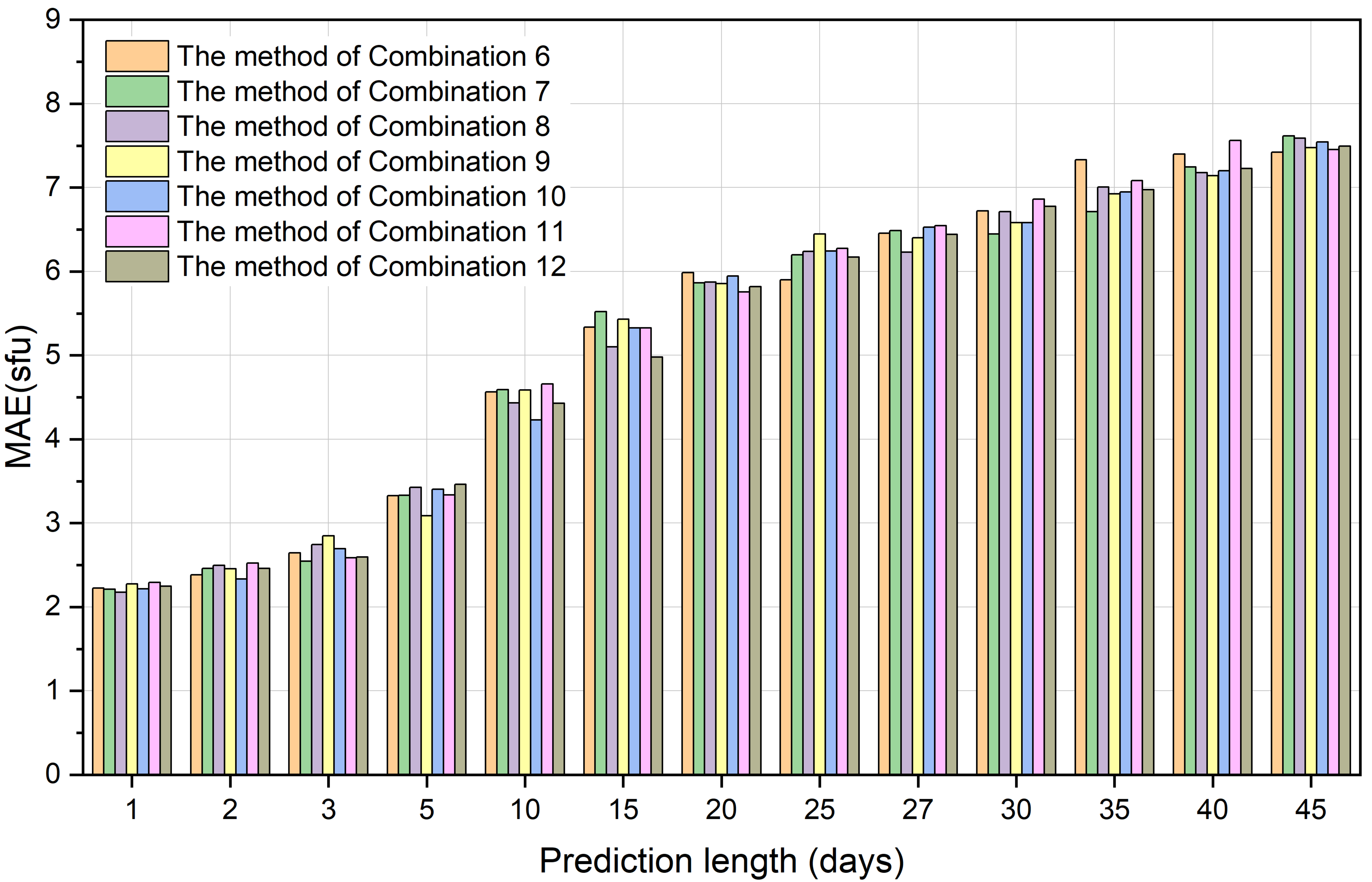}
\end{figure}
\begin{figure}[H]
	\centering
	\includegraphics[width=0.8\linewidth]{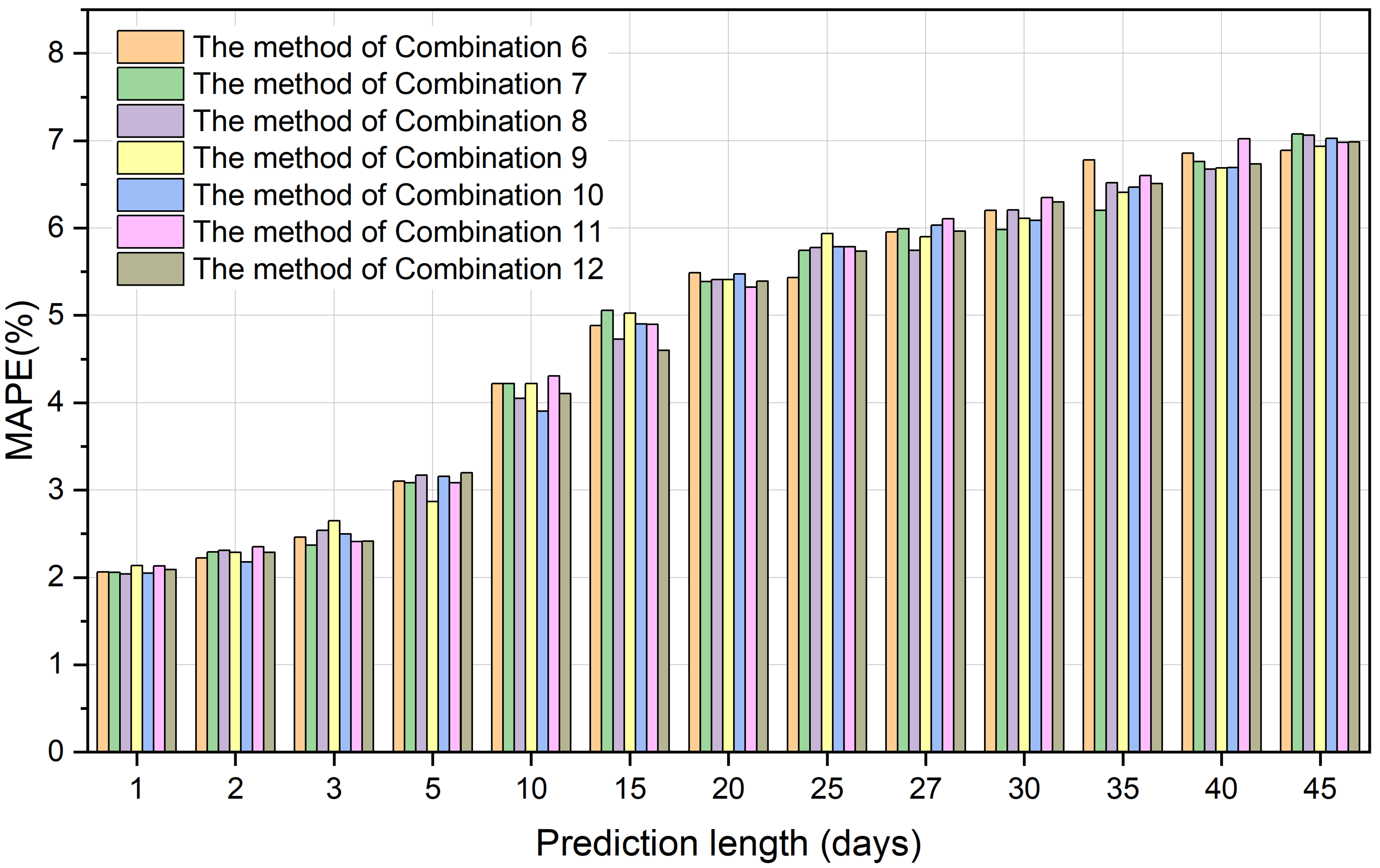}
	\caption{Bar chart comparing the prediction performance results of the Combination 6 method and the Combination 7 to 12 methods under different prediction lengths.}
	\label{fig12}
\end{figure}
\subsection{Comparison with the latest research and international institutions}\label{subsec:result4}
\begin{figure}[H]
	\centering
	\includegraphics[width=0.58\linewidth]{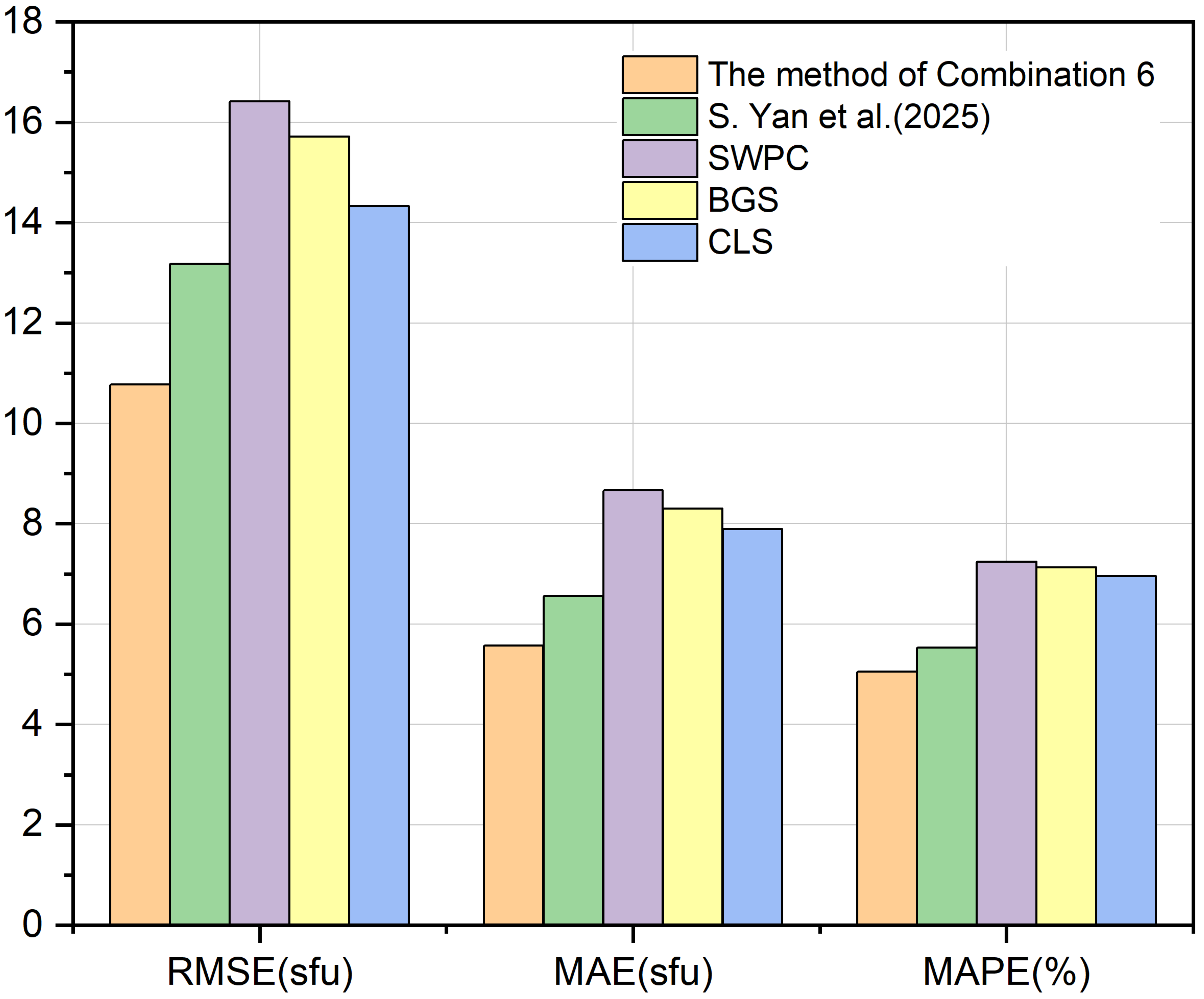}
	\caption{Bar chart of the total prediction performance of the Combination 6 method, \citet{yan2025research}, and the other three operational models.}
	\label{fig13}
\end{figure}

In this study, the Combination 6 method, which incorporates the original F10.7 index combined with its first to fifth level approximate signals and detail signals, demonstrates the optimal prediction performance in F10.7 index. To comprehensively evaluate the prediction performance superiority for the Combination 6 method, we compare its prediction performance with that of the method proposed by the latest research \citet{yan2025research} and that of the three typical operational forecast models: SWPC, BGS, and CLS. BGS and CLS predict F10.7 for future time periods daily, while SWPC releases the data for the forecast period only on Mondays since 2017. To facilitate a comparison within the same time period with the methods proposed by \citet{yan2025research} and three international institutions, we adjust the testing time period to August 3, 2018, to May 29, 2023, consistent with testing time period of \citet{yan2025research}. The forecast data from these three international institutions are available from the European Space Agency (ESA) platform (\url{https://swe.ssa.esa.int/en/forind-federated}), and the prediction performance results for the method of \citet{yan2025research} can be obtained directly from the corresponding research paper. The testing time period covers the late phase of solar cycle 24 and the early phase of solar cycle 25, during which solar flux transitioned from relatively quiet to active conditions, increasing the forecasting difficulty. Table \ref{tab7} presents the annual and total prediction performance results of the Combination 6 method, \citet{yan2025research}, and the other three operational models in the prediction length of 27 days, with bold text indicating the best performance results every column. Based on Table \ref{tab7}, Figure \ref{fig13} shows the bar chart of the overall prediction performance of the Combination 6 method, \citet{yan2025research}, and the other three operational models.

\begin{table}[!htbp]
	\centering
	\caption{The annual and total prediction performance results of the Combination 6 method, \citet{yan2025research}, and the other three operational models in the prediction length of 27 days.}
	\label{tab7}
	\setlength{\tabcolsep}{10pt}
	\begin{tabular}{ccccccccc} 
		\toprule
		Method & Evaluation Indicator & 2018 & 2019 & 2020 & 2021 & 2022 & 2023 & Total \\
		\midrule
		\multirow{3}{*}{Combination 6} 
		& RMSE & \textbf{1.357} & \textbf{1.847} & \textbf{4.649} & \textbf{7.050} & \textbf{18.735} & \textbf{23.206} & \textbf{10.774} \\
		& MAE  & \textbf{1.047} & \textbf{1.291} & \textbf{2.409} & \textbf{4.353} & \textbf{12.986}& \textbf{15.318}&\textbf{5.573} \\
		& MAPE & \textbf{1.448} & \textbf{1.827} & \textbf{2.952} & \textbf{5.288} & \textbf{9.632} & \textbf{8.771} & \textbf{5.057} \\
		\cmidrule(lr){1-9}
		\multirow{3}{*}{\citet{yan2025research}} 
		& RMSE & 1.400 & 1.998 & 4.862 & 8.714 & 21.379 & 24.278 & 13.175 \\
		& MAE  & 1.063 & 1.343 & 2.581 & 5.319 & 14.923 & 15.428 & 6.563 \\
		& MAPE & 1.526 & 1.891 & 3.176 & 5.883 & 11.063 & 9.131 & 5.531 \\
		\cmidrule(lr){1-9}
		\multirow{3}{*}{SWPC} 
		& RMSE & 1.949 & 2.387 & 6.284 & 10.670 & 25.729 & 31.746 & 16.415 \\
		& MAE  & 1.540 & 1.596 & 3.379 & 6.796 & 19.005 & 22.500 & 8.669 \\
		& MAPE & 2.212 & 2.245 & 4.116 & 7.488 & 14.270 & 13.206 & 7.243 \\
		\cmidrule(lr){1-9}
		\multirow{3}{*}{BGS} 
		& RMSE & 1.709 & 2.568 & 6.425 & 10.570 & 25.090 & 29.180 & 15.714 \\
		& MAE  & 1.285 & 1.887 & 3.605 & 6.843 & 18.143 & 19.587 & 8.299 \\
		& MAPE & 1.845 & 2.676 & 4.433 & 7.658 & 13.641 & 11.688 & 7.129 \\
		\cmidrule(lr){1-9}
		\multirow{3}{*}{CLS} 
		& RMSE & 2.468 & 2.681 & 5.936 & 9.949 & 23.072 & 25.857 & 14.326 \\
		& MAE  & 2.064 & 2.103 & 3.487 & 6.595 & 17.044 & 17.447 & 7.891 \\
		& MAPE & 3.001 & 3.012 & 4.353 & 7.400 & 12.932 & 10.512 & 6.958 \\
		\bottomrule
	\end{tabular}
\end{table}

As shown in Table \ref{tab7}, in terms of total prediction performance, the Combination 6 method achieves reductions in RMSE of 18.22\%, 34.36\%, 31.44\% and 24.79\% compared to \citet{yan2025research}, SWPC, BGS, and CLS, respectively. Corresponding reductions in MAE are 15.09\%, 35.71\%, 32.85\%, and 29.38\%, while MAPE reductions reach 8.57\%, 30.18\%, 29.06\%, and 27.32\%, respectively. As illustrated in Figure \ref{fig13}, the Combination 6 method demonstrates superior prediction performance throughout the total testing time period. Across all the values of evaluation indicators, the ranking of methods is consistent as follows: our proposed method achieves the optimal prediction performance, follow by \citet{yan2025research}, then CLS and BGS, with SWPC showing relatively lower prediction performance. These results clearly indicate that the Combination 6 method outperforms both \citet{yan2025research} and the three operational models. This might be because the statistical models used by SWPC and BGS have some limitations. These statistical models may limit the ability to capture solar activity, particularly the burst patterns during high-activity periods and the deep nonlinear relationships involved. CLS utilizes a neural network with only one hidden layer and a Logistic activation function. While this relatively simple structure is suitable for processing simple datasets, its model capacity and expressive power may be insufficient to capture the deep-level complex associations and long-term dependencies inherent in the F10.7 index. In order to effectively distinguish the basic trends and transient disturbances of solar activity, \citet{yan2025research} adopted a factor decomposition method. This method did not focus on the intrinsic relationship between the trend components and the disturbance components during the training process, and used an SG filter for a single decomposition. We adopt the method based on wavelet decomposition, which captures the intrinsic relationship between the approximate signals that reflect the long-term trend and the detail signals that represent short-term disturbance through end-to-end overall modeling. Our method also utilizes wavelet-decomposed signals of the first to fifth levels, providing the model with more comprehensive feature inputs. Meanwhile, the iTransformer model we adopt effectively captures the deep relationship between wavelet-decomposed signals of different levels and the F10.7 index, enabling the model to fully utilize the feature information they provide. Therefore, the method based on wavelet decomposition outperforms the method based on factor decomposition in terms of prediction performance.

To further evaluate the superiority of the Combination 6 method, we divide the data throughout the testing time period into four solar activity conditions based on the numerical range of the F10.7 index \citep{licata2020benchmarking,YE20246309}. This comparison allows for a clear demonstration of the prediction performance differences among various methods under different conditions, such as solar minimum and maximum. Table \ref{tab8} shows the prediction performance results of the Combination 6 method and the three operational models across four conditions in the prediction length of 27 days, with bold text indicating the best performance results under different conditions.

\begin{table}[H]
	\centering
	\caption{Prediction performance of the Combination 6 method and the three operational models across four conditions in the prediction length of 27 days.}
	\label{tab8}
	\setlength{\tabcolsep}{10pt}
	\begin{tabular}{ccccc}
		\toprule
		Condition & Model & RMSE & MAE & MAPE \\
		\midrule
		\multirow{4}{*}{Low (F10.7 $\leq$ 75)} 
		& Combination 6 & \textbf{2.121} & \textbf{1.459} & \textbf{2.062} \\
		& BGS & 2.748 & 1.929 & 2.728 \\
		& CLS & 2.977 & 2.219 & 3.165 \\
		& SWPC & 2.470 & 1.726 & 2.437 \\
		\midrule
		\multirow{4}{*}{Moderate (75 $<$ F10.7 $\leq$ 150)} 
		& Combination 6 & \textbf{11.836} &\textbf{8.696} & \textbf{7.807} \\
		& BGS & 16.521 & 11.837 & 10.878 \\
		& CLS & 15.082 & 11.292 & 10.380 \\
		& SWPC & 16.043 & 11.950 & 10.905 \\
		\midrule
		\multirow{4}{*}{Elevated (150 $<$ F10.7 $\leq$ 190)} 
		& Combination 6 & \textbf{18.237} & \textbf{14.288} & \textbf{8.735} \\
		& BGS & 22.831 & 18.245 & 11.159 \\
		& CLS & 19.701 & 15.928 & 9.766 \\
		& SWPC & 25.983 & 22.442 & 13.707 \\
		\midrule
		\multirow{4}{*}{High ( F10.7 $>$ 190)} 
		& Combination 6 & \textbf{52.215} & \textbf{39.767} & \textbf{17.113} \\
		& BGS & 72.714 & 59.614 & 25.930 \\
		& CLS & 66.740 & 51.945 & 22.289 \\
		& SWPC & 71.618 & 61.669 & 27.057 \\
		\midrule
		\multirow{4}{*}{Total ($-\infty < \text{F10.7} < +\infty$)} 
		& Combination 6 & \textbf{10.774}& \textbf{5.573} & \textbf{5.057} \\
		& BGS & 15.714 & 8.299 & 7.129 \\
		& CLS & 14.326 & 7.891 & 6.958 \\
		& SWPC & 16.415 & 8.669 & 7.243 \\
		\bottomrule
	\end{tabular}
\end{table}

As shown in Table \ref{tab8}, in the low activity condition, the Combination 6 method demonstrates significantly superior prediction performance across all evaluation indicators compared to the other models. In the moderate activity condition, compared to CLS, the best performing model among the three operational models, the Combination 6 method achieves reductions of 21.53\% in RMSE, 22.99\% in MAE, and 24.79\% in MAPE. In the elevated activity condition, the Combination 6 method reduces RMSE, MAE, and MAPE by 7.43\%, 10.30\%, and 10.56\%, respectively, compared to CLS, which again shows the best performance among the three operational models. In the high activity condition, the Combination 6 method demonstrates outstanding prediction performance. Its RMSE, MAE and MAPE are significantly lower than those of the best performing CLS among the three operational models, with reductions of up to 21.76\%, 23.44\% and 23.22\%, respectively. These results indicate that the Combination 6 method can more effectively predict the F10.7 index across different conditions, demonstrating superior prediction performance under various solar activity conditions and significantly outperforming the three operational models. The Combination 6 method demonstrates outstanding prediction performance under both high and low activity conditions. This is likely because during high activity condition, the F10.7 index exhibits abrupt spikes, and wavelet decomposition can effectively extract the detail signals representing transient fluctuations, thereby enabling the model to more accurately capture the intricate details of these sharp variations. Under low solar activity condition, where the F10.7 index varies smoothly, wavelet decomposition helps extract approximate signals representing long-term trends, thereby improving the ability of model to depict the overall gradual evolution. Furthermore, the iTransformer model we employ fully captures the multi-scale relationships between the wavelet-decomposed signals and the F10.7 index. As a result, the Combination 6 method not only excels in high and low activity conditions but also shows strong prediction performance across other activity conditions and overall condition.

\subsection{Generalization performance on Dataset B}\label{subsec:result5}
To validate whether the Combination 6 method proposed in this study maintains excellent prediction performance on different telescope observation data within the same time period, we conduct the generalization performance test on Dataset B. During the testing process, we transfer the Combination 6 method to the PatchTST model used by \citet{YE20246309} and compare its prediction performance with that of the method of \citet{YE20246309}. It is worth noting that the method of \citet{YE20246309} was only tested on Langfang data and not evaluated on DRAO data from the same time period. Therefore, we retrain the method of \citet{YE20246309} on our dataset and test it on Dataset B. Tables \ref{tab9} and \ref{tab10} present the prediction performance results of the Combination 6 method and the method of \citet{YE20246309} on Dataset B under different forecast lengths, where bold text indicates the best prediction performance results for different prediction lengths. The visualization results of Tables \ref{tab9} and \ref{tab10} are shown in Figure \ref{fig14}.

\begin{figure}[H]
	\centering
	\includegraphics[width=0.6\linewidth]{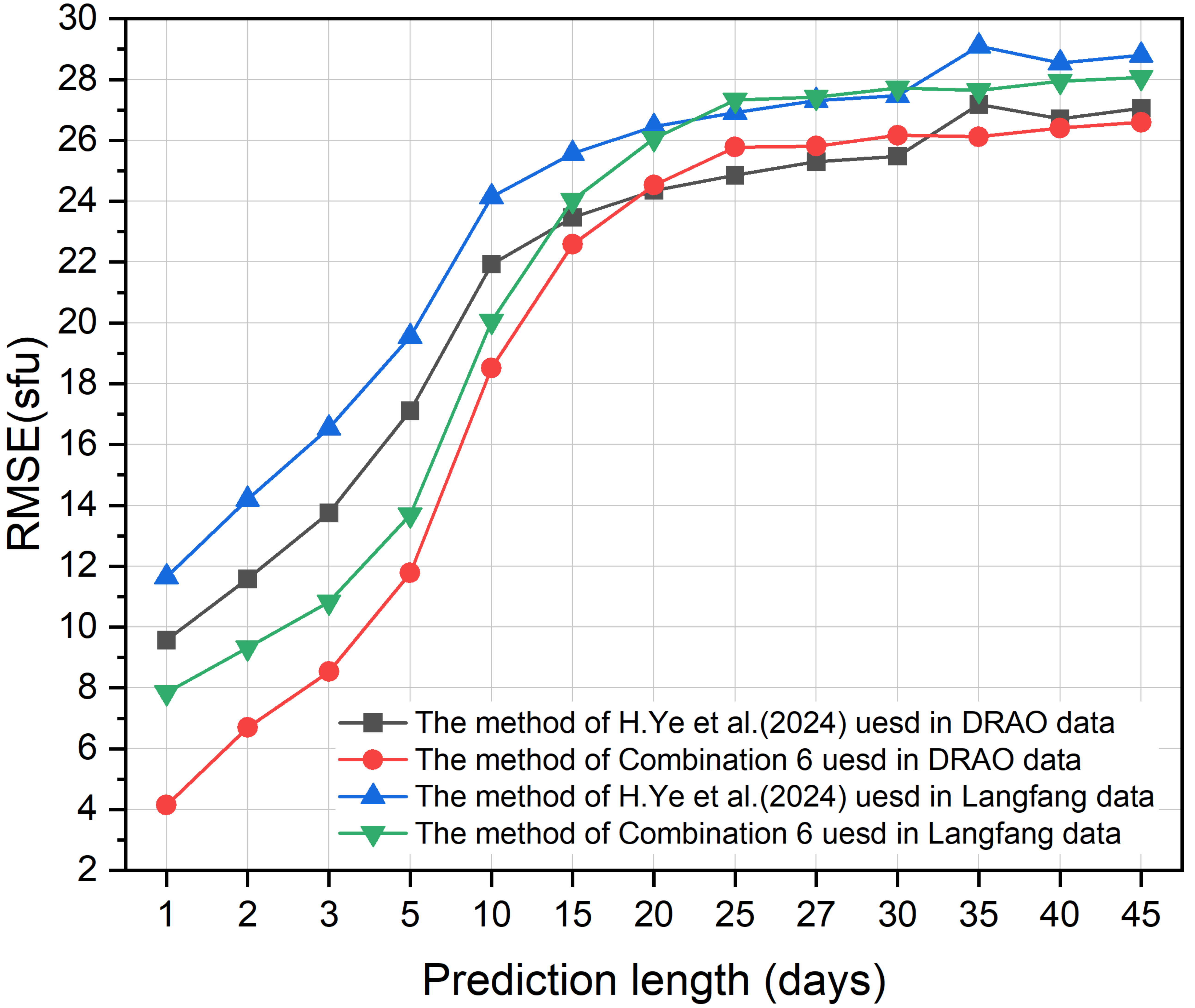}
\end{figure}
\begin{figure}[H]
	\centering
	\includegraphics[width=0.6\linewidth]{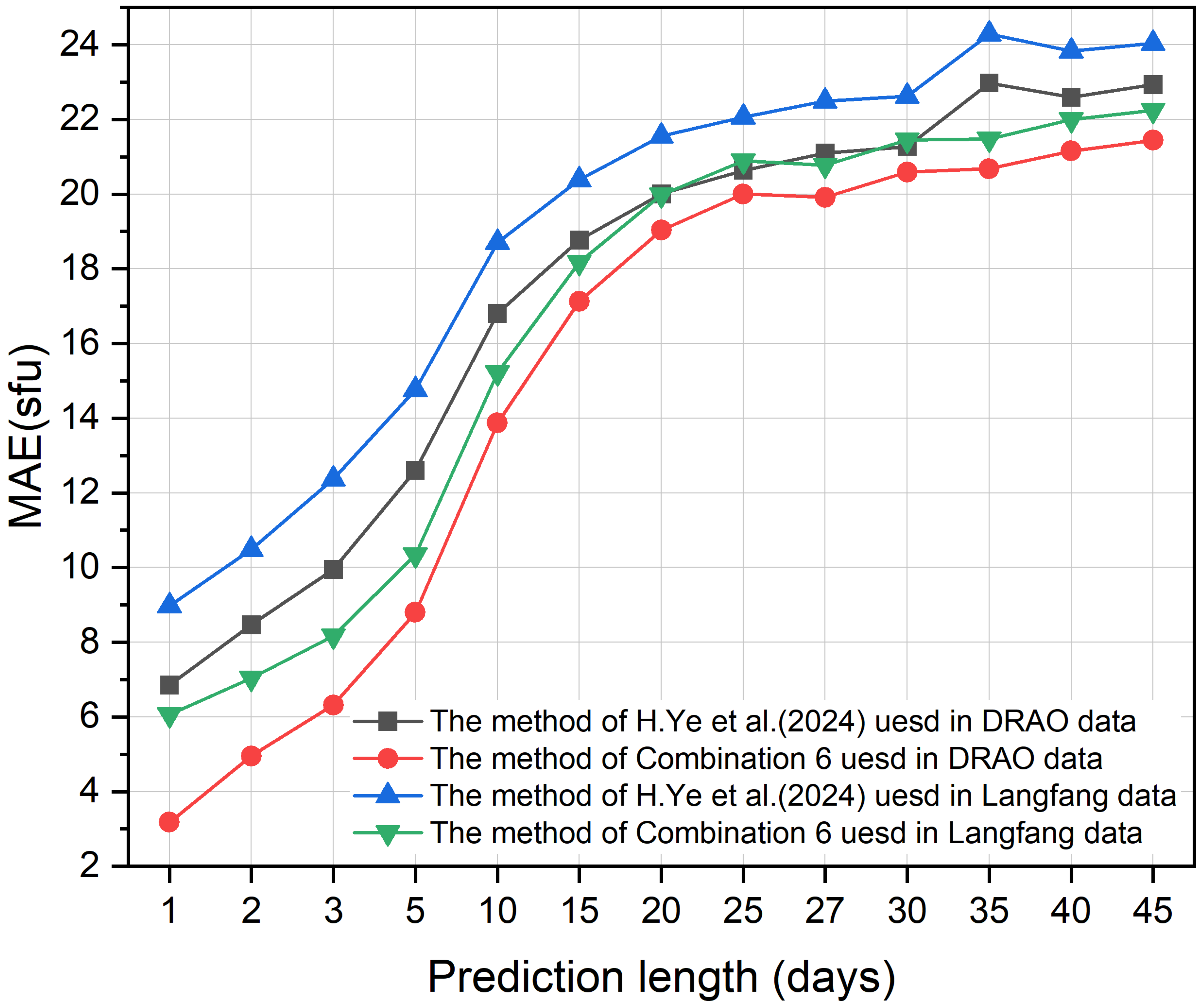}
\end{figure}
\begin{figure}[H]
	\centering
	\includegraphics[width=0.6\linewidth]{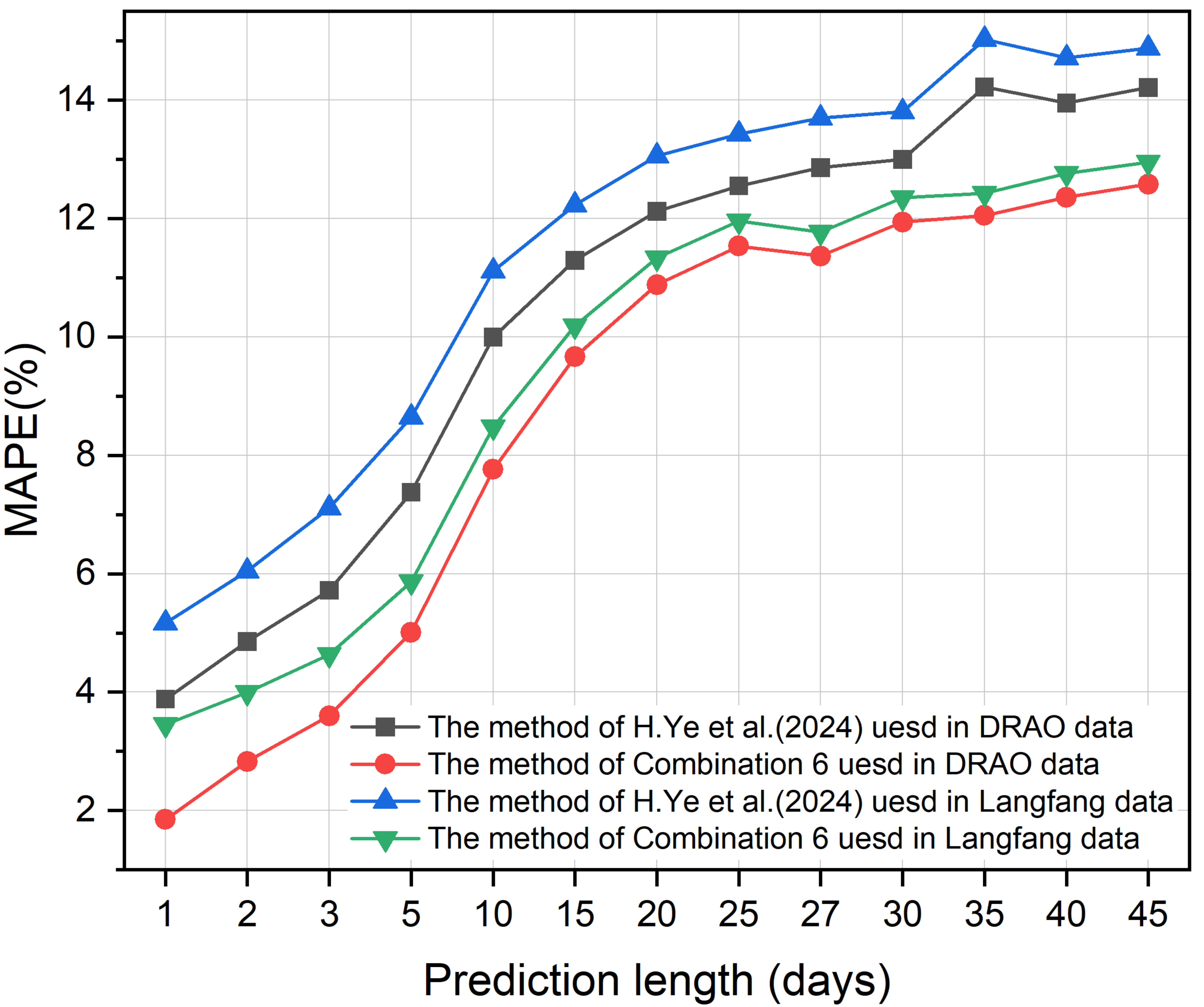}
	\caption{On the DRAO data and Langfang data, the line graphs showing the prediction performance of the Combination 6 method and the method of \citet{YE20246309} under different prediction lengths.}
	\label{fig14}
\end{figure}

\begin{table}[H]
	\centering
	\caption{Prediction performance results of the Combination 6 method and the method of \citet{YE20246309} on the DRAO data under different forecast lengths.}
	\setlength{\tabcolsep}{10pt}
	\begin{tabular}{ccccccc}
		\toprule[1pt]
		\multirow{2}{*}{Prediction length (days)}& \multicolumn{3}{c}{The method of \citet{YE20246309}} & \multicolumn{3}{c}{Combination 6} \\
		\cmidrule{2-7}
		& RMSE   & MAE    & MAPE   & RMSE   & MAE    & MAPE \\
		\midrule
		1     & 9.568  & 6.854  & 3.884  & \textbf{4.144} & \textbf{3.176} & \textbf{1.850} \\
		2     & 11.582 & 8.466  & 4.855  & \textbf{6.698} & \textbf{4.956} & \textbf{2.829} \\
		3     & 13.748 & 9.947  & 5.719  & \textbf{8.529} & \textbf{6.320} & \textbf{3.601} \\
		5     & 17.108 & 12.600 & 7.377  & \textbf{11.783} & \textbf{8.795} & \textbf{5.010} \\
		10    & 21.933 & 16.807 & 9.996  & \textbf{18.515} & \textbf{13.870} & \textbf{7.764} \\
		15    & 23.466 & 18.77  & 11.294 & \textbf{22.577} & \textbf{17.127} & \textbf{9.663} \\
		20    & \textbf{24.352} & 20.005 & 12.120 & 24.530 & \textbf{19.034} & \textbf{10.882} \\
		25    & \textbf{24.852} & 20.633 & 12.552 & 25.776 & \textbf{20.006} & \textbf{11.536} \\
		27    & \textbf{25.295} & 21.098 & 12.860 & 25.814 & \textbf{19.916} & \textbf{11.366} \\
		30    & \textbf{25.476} & 21.273 & 12.998 & 26.167 & \textbf{20.587} & \textbf{11.940} \\
		35    & 27.179 & 22.979 & 14.221 & \textbf{26.119} & \textbf{20.683} & \textbf{12.052} \\
		40    & 26.709 & 22.596 & 13.950 & \textbf{26.406} & \textbf{21.154} & \textbf{12.357} \\
		45    & 27.065 & 22.934 & 14.212 & \textbf{26.597} & \textbf{21.445} & \textbf{12.580} \\
		\bottomrule[1pt]
	\end{tabular}
	\label{tab9}
\end{table}

\begin{table}[H]
	\centering
	\caption{Prediction performance results of the Combination 6 method and the method of \citet{YE20246309} on the Langfang data under different forecast lengths.}
	\setlength{\tabcolsep}{10pt}
	\begin{tabular}{ccccccc}
		\toprule[1pt]
		\multirow{2}{*}{Prediction length (days)}& \multicolumn{3}{c}{The method of \citet{YE20246309}} & \multicolumn{3}{c}{Combination 6} \\
		\cmidrule{2-7}
		& RMSE   & MAE    & MAPE   & RMSE   & MAE    & MAPE \\
		\midrule
		1     & 11.645	 & 8.978 & 	5.167	 &\textbf{ 7.851}	 & \textbf{6.065} & 	\textbf{3.457} \\
		2     & 14.202 & 10.488	 & 6.046 & 	\textbf{9.322} & 	\textbf{7.037} & 	\textbf{3.997} \\
		3     & 16.535 & 	12.367	 & 7.116 & 	\textbf{10.824} & 	\textbf{8.16}4 & 	\textbf{4.635} \\
		5     & 19.551 & 	14.766 & 	8.646 & 	\textbf{13.693} & 	\textbf{10.343}	 & \textbf{5.867} \\
		10    & 24.138	 & 18.712 & 	11.113 & 	\textbf{20.058} & 	\textbf{15.225} & 	\textbf{8.479} \\
		15    & 25.570	 & 20.385 & 	12.226 & 	\textbf{24.037}	 & \textbf{18.175} & 	\textbf{10.185} \\
		20    & 26.462 & 	21.561 & 	13.060 & 	\textbf{26.065} & 	\textbf{19.982} & 	\textbf{11.337} \\
		25    & \textbf{26.914} & 	22.069 & 	13.429	 & 27.328	 & \textbf{20.897}	 & \textbf{11.958} \\
		27    & \textbf{27.312} & 	22.487 & 	13.697	 & 27.423	 & \textbf{20.781} & 	\textbf{11.773} \\
		30    & \textbf{27.466} & 	22.629 & 	13.805 & 	27.730	 &\textbf{21.449} & 	\textbf{12.356} \\
		35    & 29.097 & 	24.294	 & 15.026 & 	\textbf{27.646} & 	\textbf{21.476 }& 	\textbf{12.429} \\
		40    & 28.550 & 	23.831	 & 14.715 & 	\textbf{27.948} & 	\textbf{21.996} & 	\textbf{12.762} \\
		45    & 28.796 & 	24.043	 & 14.879 & 	\textbf{28.074}	 &\textbf{ 22.247} & 	\textbf{12.953} \\
		\bottomrule[1pt]
	\end{tabular}
	\label{tab10}
\end{table}

Based on the performance results from Tables \ref{tab9}, \ref{tab10}, and Figure \ref{fig14}, the Combination 6 method shows similar trends and numerical ranges in RMSE, MAE, and MAPE on the both Langfang and DRAO data. Compared to the performance differences of the method of \citet{YE20246309} between the Langfang and DRAO data, the Combination 6 method exhibits comparable performance differences in RMSE between these two data, but significantly smaller differences in MAE and MAPE. This indicates that the Combination 6 method demonstrates superior generalization performance. Across all prediction lengths, the prediction performance of the Combination 6 method outperforms that of the method of \citet{YE20246309}. These results demonstrate that the Combination 6 method possesses better generalization performance and superior prediction performance. This advantage may stem from the combination of multiple approximate and detail signals providing the model with richer and more comprehensive feature information, enabling it to capture both the macroscopic patterns and microscopic details of solar activity simultaneously. Our method reduces dependence on single feature variable and enhances the adaptability of model to different data, thereby improving the generalization performance of model and boosting the prediction performance of F10.7 index.

\section{Conclusions and discussions}\label{sec:Conclusions} 
In this study, we construct Dataset A based on DRAO data for model training, validation, and testing, and Dataset B based on DRAO and Chinese Langfang data for generalization performance testing. We are the first to propose a novel F10.7 forecasting method based on the wavelet decomposition. Through wavelet decomposition, we obtain approximate signals and detail signals of the F10.7 index, which are used together with the original F10.7 index as inputs to the iTransformer model to jointly predict future F10.7 index. Subsequently, we introduce the ISN which is closely related to the F10.7 index, and its approximate and detail signals, to investigate their impact on prediction performance. Next, we compare the prediction performance of our optimal method with that of \citet{yan2025research} and three operational models from international institutions (SWPC, BGS, CLS) within the same time period. Additionally, we conduct generalization performance study on Dataset B by transferring the Combination 6 method to the PatchTST model used by \citet{YE20246309} and comparing our method with the method of \citet{YE20246309}.

The main results of this study are summarized as follows: (1) On Dataset A, compared to the baseline method using only the original F10.7 index, the wavelet-based combination (Combinations 2 to 6) methods demonstrate improved prediction performance. The incremental incorporation of higher-level approximate and detail signals leads to a improvement in prediction performance. Among them, the Combination 6 method, which incorporates the original F10.7 index combined with its first to fifth level approximate signals and detail signals, exhibits the best prediction performance. The Combination 6 method demonstrates superior prediction performance compared to both the Combination A method and Combination D method. The former incorporates the original F10.7 index with approximate signals from the first to fifth levels, while the latter incorporates the original F10.7 index with detail signals from the same levels. (2) On Dataset A, the method of adding ISN and its wavelet decomposition signals to Combination 6 does not improve the prediction performance. (3) The Combination 6 method demonstrates superior prediction performance compared to \citet{yan2025research} and three operational models. In terms of total prediction performance, the Combination 6 method achieves reductions in RMSE of 18.22\%, 34.36\%, 31.44\% and 24.79\% compared to \citet{yan2025research}, SWPC, BGS, and CLS, respectively. Corresponding reductions in MAE are 15.09\%, 35.71\%, 32.85\%, and 29.38\%, while MAPE reductions reach 8.57\%, 30.18\%, 29.06\%, and 27.32\%. Furthermore, the Combination 6 method exhibits excellent prediction performance of F10.7 index across different solar activity conditions and outperforms the three operational models. (4) On Dataset B, the Combination 6 method demonstrates superior generalization performance and better prediction performance across all forecast lengths compared to the method of \citet{YE20246309}.

This study significantly enhances the prediction performance of the F10.7 index by introducing wavelet multi-scale decomposition. Compared to the baseline combination method using only the original F10.7 index, the prediction performance of the combination methods based on wavelet multi-scale decomposition shows an overall upward trend. This indicates that incorporating approximate signals and detail signals obtained through wavelet decomposition can effectively enhance the ability of model to predict F10.7 index. Among them, the Combination 6 method achieves the best prediction performance, with its advantages potentially stemming from two aspects. On one hand, this combination integrates wavelet-decomposed signals, which contain comprehensive features from low level to high level. This enables the model to learn feature information from the data more fully. On the other hand, compared to the Combination A method using only approximate signals, and the Combination D method using only detail signals, the Combination 6 method achieves a coordinated modeling of the overall patterns and local disturbances of solar activity by simultaneously integrating long-term trends and detail fluctuations. Compared with the factor decomposition method adopted by \citet{yan2025research}, our overall modeling method retains the synergistic effect between the approximate signals and the detail signals, and provides the model with more abundant multi-scale feature inputs, thereby achieving better prediction performance. We apply the wavelet decomposition method to the PatchTST model used by \citet{YE20246309}. Compared with the single-variable prediction method used by \citet{YE20246309}, our method achieves better generalization performance and prediction performance by utilizing the multi-level approximate signals and detail signals.

In the future, we will introduce more signal decomposition methods, such as VMD, singular spectrum analysis, and EMD \citep{liu2018smart,zhang2021hybrid,wang2014forecasting,hao2024f10}, for comparative studies to further enhance the prediction performance of the model. Simultaneously, we will incorporate multi-source F10.7 observations from different telescopes and more feature data highly correlated with F10.7 to investigate their impact on model prediction performance. Additionally, we plan to develop a real-time, high-precision forecasting platform to predict the F10.7 index for the future time periods, providing data services for space weather warning, satellite communication, and other related fields.
\section*{Acknowledgments}
We sincerely thank the anonymous reviewers for their constructive comments, which have significantly improved the quality of this work. We also gratefully acknowledge the Dominion Radio Astrophysical Observatory (DRAO), the British Geological Survey (BGS), Collect Localization Satellites (CLS), the Space Weather Prediction Center (SWPC), and the Long and Short Wave Solar Precision Flux Radio Telescope (L\&S) in Langfang, China, for providing the essential data that formed the foundation of this research. This research is supported by the National Natural Science Foundation of China (Grant No. 12473056) and the Jiangsu Province Natural Science Foundation (Grant No. BK20241830) and Qing Lan Project.
\section*{Data availability}
The data and codes used in this study are available via doi: (\url{10.5281/zenodo.17777755}).
\bibliography{sample701}{}
\bibliographystyle{aasjournalv7}

\end{document}